%% file: arxiv.tex
\documentclass[11pt,draftcls,onecolumn]{IEEEtran}

\usepackage{makeidx}
\usepackage{amsmath}
\usepackage{epsfig}
\usepackage{amssymb}
\usepackage{textcomp}
\usepackage{pstricks,pst-text}
\usepackage{algorithm}
\usepackage{algorithmic}
\usepackage{datetime}
\usepackage{setspace}
\usepackage{watermark}
\usepackage{graphicx}
\usepackage{caption}
\usepackage{subcaption}
\usepackage{cases}
\input{core/macros.tex}

\newcommand{\firstAuthor}       {Raj~Thilak~Rajan,~\IEEEmembership{Student Member,~IEEE}}
\newcommand{\secondAuthor}      {Geert~Leus,~\IEEEmembership{Fellow,~IEEE}}
\newcommand{\thirdAuthor}       {Alle-Jan~van~der~Veen,~\IEEEmembership{Fellow,~IEEE}}
\newcommand{\theTitle}          {Joint relative position and velocity estimation \\ for an anchorless network of mobile nodes}

\newcommand{\dr}    {\dot{r}}
\newcommand{\ddr}   {\ddot{r}}
\newcommand{\dddr}  {\dddot{r}}
\newcommand{\bdr}   {\dot{\br}}
\newcommand{\bddr}  {\ddot{\br}}

\newcommand{\dur}   {\dot{\ur}}
\newcommand{\ddur}  {\ddot{\ur}}

\newcommand{\bdur}  {\dot{\bur}}
\newcommand{\bddur} {\ddot{\bur}}

\newtheorem{theorem}{Theorem}

\newcounter{remarkCounter}
\stepcounter{remarkCounter}

\title  {\theTitle}
\author {
\firstAuthor\ , \\ \secondAuthor\ and \thirdAuthor
\thanks{R.T.Rajan is with Netherlands Institute for Radio Astronomy (ASTRON), Dwingeloo, The Netherlands (email: rajan@astron.nl) and TU Delft, Delft, The Netherlands.}
\thanks{Geert Leus and A.-J. van der Veen are with TU Delft, Delft, The Netherlands (email: g.j.t.leus@tudelft.nl; a.j.vanderveen@tudelft.nl)}
\thanks{This research was funded in part by the STW OLFAR project (Contract Number: 10556) within the ASSYS perspectief program.}
}

\begin{document}
\watermark{\small{R.T.Rajan, G. Leus, A.-J.van der Veen}}
%\watermark{\small{To be submitted to \emph{IEEE} Transactions on Signal Processing}}
\maketitle

\begin{abstract} Localization is a fundamental challenge for any wireless network of nodes, in particular when the nodes are mobile. We present an extension of the classical Multidimensional scaling (MDS) for an \emph{anchorless network of mobile nodes}, wherein the solutions to the time-varying relative node positions are shown to lie in the derivatives of the time-varying inter-nodal pairwise distances. Moreover, we show that the relative position of a mobile node at each time instance is only dependent on the initial relative position, relative velocity and a common rotation matrix of the respective node, which are estimated using MDS-like and least squares estimators. Simulations are conducted to evaluate the performance of the proposed solutions and the results are presented.

\IEEEkeywords\ relative position and velocity, rotation matrix, Multi-Dimensional Scaling (MDS), dynamic ranging, anchor-free wireless network, \Cramer\ Rao Bounds
\end{abstract}

\section{Introduction}Localization is a key requirement for the deployment of wireless networks in a wide range of applications. There are numerous absolute localization algorithms, such as Time of Arrival (ToA), Time Difference of Arrival (TDoA) and Received Signal Strength (RSS) which cater to anchored networks, where only the positions of a few nodes are known \cite{patwari05}. Alternatively, when there are no reference anchors, then the relative positions of the nodes, up to a rotation and translation, can still be obtained using Multi-Dimensional Scaling (MDS) based solutions \cite{borg97,cheung05tsp}. Such anchorless networks arise naturally when the nodes are deployed in inaccessible locations or when anchor information is known intermittently. In both anchored or anchorless scenarios, pairwise distances are one of the key inputs for almost all localization techniques. For stationary nodes, these pairwise distances are classically obtained by measuring the propagation delays of multiple time stamp exchanges between the nodes and averaging these measurements over a time period.

A step further, when the nodes are mobile, then conventionally either the nodes are considered relatively stationary within desired accuracies for the complete duration of the measurement interval (\ie multiple distance measurements) \cite{rajan2011camsap} or Doppler measurements are utilized \cite{wei10}. Unfortunately, Doppler measurements are not always available and the assumption on the node positional stability for large time periods is not necessarily practical. For a mobile network, the application of classical MDS-based relative positioning at every time instant yields a sequence of position matrices with arbitrary rotation, thereby providing no information on the relative velocities of the nodes. The term \emph{relative velocities} indicates the velocity vectors of the nodes, up to a common rotation, translation and reflection. To the best of the authors' knowledge, the estimation of relative velocities for an anchorless network has not yet been investigated in literature.

\subsection{Applications} Our motivation for this work is triggered by \emph{inaccessible} mobile wireless networks, which have partial or no information of absolute coordinates and/or clock references. Such scenarios are prevalent in under-water communications \cite{chandrasekhar2006}, indoor positioning systems \cite{liu2007} and envisioned space based satellite networks with minimal ground segment capability. A particular project of interest is Orbiting Low Frequency Antennas for Radio astronomy (OLFAR) \cite{rajanIEEEAero11}, a Dutch funded program which aims to design and develop a detailed system concept for a scalable interferometric array of more than ten identical, autonomous satellites in space (far from earth) to be used as a scientific instrument for ultra low frequency observations ($0.3$ kHz - $30$ MHz). Due to limitations of earth-based tracking, the OLFAR cluster will be an independent cooperative network of nodes, whose positions and velocities need to be estimated jointly.

\subsection{Contributions} In this article, our quest is to understand the relative kinematics of an \emph{anchorless network of mobile} nodes, with or without any information on the Doppler measurements. By the term anchorless, we emphasize that the absolute positions and the velocities of the nodes are unknown.  We begin by approximating the time-varying pairwise propagation delays (and subsequently the ranges) between the mobile nodes as a Taylor series in time, which is aptly termed dynamic ranging (Section \ref{sec:dynamicRanging}). A simple yet efficient time based monomial basis is employed, to estimate the derivatives of the pairwise distances at a given time instant (Section \ref{sec:dynamicRangingAlgorithm}). Under the assumption of constant velocity for a short duration of time, we show that the relative position of each node is dependent only on the initial relative position, the relative velocity and a unique rotation matrix (Section \ref{sec:relativeKinematics}). Furthermore, the solutions to the unknown initial relative position, the relative velocity and the rotation matrix lie in the first three derivatives of the time-varying pairwise distance. Subsequently, we present a MDS-like and least squares solutions to estimate the unknown parameters in Section \ref{sec:algorithms} and \Cramer\ Rao Bounds are derived. Simulations are conducted to evaluate the performance of the dynamic ranging algorithm and the MDS based estimators for relative Positions and Velocities (Section \ref{sec:simulations}).

\textit{Notation:} The element wise matrix Hadamard product is denoted by $\odot$, $(\cdot)^{\odot N}$ denotes element-wise matrix
exponent and $\oslash$ indicates the element-wise Hadamard division. The Kronecker product is indicated by $\otimes$ and the transpose operator by ($\cdot)^T$. $\b1_N = [1, 1 \hdots ,1]^T, \bzero_N = [0, 0 \hdots, 0]^T \in \mathbb{R}^{N \times 1}$, are vectors of ones and zeros, respectively. The Euclidean norm is denoted by $\norm{\cdot}$, $\bI_N$ is a $N \times N$ identity matrix and $\bzero_{M,N}$ is a $M \times N$ matrix of zeros. A diagonal matrix of the vector $\ba$ is represented by $\diag(\ba)$ and a block diagonal matrix $ \bA= \bdiag (\bA_1, \bA_2, \hdots, \bA_N)$ consists of matrices $\bA_1, \bA_2, \hdots, \bA_N$ along the diagonal and $0$ elsewhere. $\text{vec}(\bA)$ operator reshapes the matrix $\bA$ into a vector. $a \sim \cN (\mu,\Sigma)$ is shorthand for a randomly distributed Gaussian variable with mean $\mu$ and variance $\Sigma$.

\section{Dynamic ranging} \label{sec:dynamicRanging} \subsection{Range model}
Consider a cluster of $N$ nodes in a $P$-dimensional Euclidean space. If the nodes are fixed, then the pairwise propagation delay at time $t_0$ between a given node pair $(i,j)$ is defined as \begin{eqnarray} \tau_{ij} (t_0) \equiv\ \tau_{ji}(t_0) \triangleq c^{-1}d_{ij}(t_0), \end{eqnarray} where $d_{ij}(t_0)$ is the fixed distance between the node pair at $t_0$ and $c$ is the speed of the electromagnetic wave in the medium. However, when the nodes are mobile, the relative distances between the nodes are a non-linear function of time (for $P \ge 2$), even when the nodes are in linear motion\footnote{Later in the article, we will assume the nodes to be in constant velocities. However, here we present a generalized Taylor approximation of the time-varying pairwise distance, for any motion.}. For a small time interval $\Delta t = t - t_0$, we consider these relative distances as a smoothly varying polynomial. The propagation delay $\tau_{ij}(t) \equiv \tau_{ji}(t)$ between a given node pair $(i,j)$ is then (classically) an infinite Taylor series around a time instant $t_0$ within the neighborhood $\Delta t$. As an extension of the linear range model \cite{rajan2012icassp}, we have  \begin{eqnarray}
\tau_{ij}(t_0 +\Delta t) \triangleq c^{-1}d_{ij}(t_0 +\Delta t)  \triangleq c^{-1}d_{ij}(t),
\label{eq:propagationDefinition} \end{eqnarray} where $d_{ij}(t)$ is the distance at $t= t_0 +\Delta t$, given by \begin{eqnarray}
d_{ij}(t) &=& r_{ij} + \frac{\dr_{ij}}{1!}\Delta t + \frac{\ddr_{ij}}{2!}\Delta t^2 + \hdots,
\label{eq:rangeDefinition}
\end{eqnarray} where $\bthetau_{ij}=[r_{ij}, \dr_{ij},\ddr_{ij},  \hdots ] \in \mathbb{R}^{L \times 1}$ are the range parameters. The first coefficient $r_{ij} \equiv\ d_{ij}(t_0)$ is the initial pairwise distance and the following $L-1$ coefficients are successive derivatives of $r_{ij}$ at $t_0$. Without loss of generality, assuming $t_0 =0$, we have $t = \Delta t$ and subsequently (\ref{eq:propagationDefinition}) and (\ref{eq:rangeDefinition}) simplify to the Maclaurian series as \begin{eqnarray}
\tau_{ij}(t)  =  c^{-1} \Big(r_{ij} + \dr_{ij}t + \frac{\ddr_{ij}}{2!}t^2  + \hdots\Big).
\label{eq:rangeDefinition1} \end{eqnarray} The \emph{unique} pairwise ranges between all the $N$ nodes are collected in a vector $\br \in \mathbb{R}^{\bar{N} \times 1}$, where $\bar{N}= \begin{pmatrix} N \\ 2 \end{pmatrix}$ is the number of unique pairwise baselines. Along similar lines, we can define $\bdr \in \mathbb{R}^{\bar{N} \times 1}$, $\bddr \in \mathbb{R}^{\bar{N} \times 1}$ and corresponding higher-order terms. The polynomial range basis is simplified further by introducing  \begin{equation}
%\begin{bmatrix} \ur_{ij}, \dur_{ij}, \ddur_{ij},  \hdots \end{bmatrix} = c^{-1} \begin{bmatrix} r_{ij}, \dfrac{\dr_{ij}}{1!}, \dfrac{\ddr_{ij}}{2!}, \hdots \end{bmatrix}
\begin{bmatrix} \ur_{ij}, \dur_{ij}, \ddur_{ij},  \hdots \end{bmatrix}^T = \diag(\bff) ^{-1} \begin{bmatrix} r_{ij}, \dr_{ij}, \ddr_{ij}, \hdots \end{bmatrix}^T
\label{eq:rangeTranslation}
\end{equation}  where $\bff= c[1,\ 1!,\ 2!,\ \hdots]^T \in \mathbb{R}^{L \times 1}$, such that  (\ref{eq:rangeDefinition1}) is \begin{eqnarray}
\tau_{ij}(t) = c^{-1} d_{ij}(t)  &\triangleq& \ur_{ij} + \dur_{ij}t + \ddur_{ij}t^2  + \hdots
\label{eq:rangeDefinition2} \end{eqnarray} Following the definition of $\btheta= \begin{bmatrix} \br, \bdr, \bddr, \hdots \end{bmatrix}$, we define $\bur \in \mathbb{R}^{\bar{N} \times 1}$, $\bdur \in \mathbb{R}^{\bar{N} \times 1}$, $\bddur \in \mathbb{R}^{\bar{N} \times 1}$ and similarly higher-order terms.

\textit{{\bf Remark \arabic{remarkCounter}}: (Doppler measurements): Observe that in essence, $\br$ is the ToA at $t_0$, the range rate $\bdr$ is the radial velocity (as obtained from a Doppler shift) and the second order range parameter $\bddr$ is the rate of radial velocity (as observed from a Doppler spread) between the nodes at $t=t_0$. These range coefficients can be readily incorporated if these measurements are available.} \stepcounter{remarkCounter}

\begin{figure}[tp] \centering

%%% Double colomn
%\includegraphics[scale=0.30]{figs/dia/pairwise.eps}
%\caption{Communication between a pair of mobile nodes where
%the nodes transmit and receive, during which $K$ time stamps are recorded at respective nodes.
%Similar to \cite{rajan2012icassp,rajanTSP1}, we put no pre-requisite on the sequence, direction or number of communications.}
%\rput(-3.9,5.7) {\scriptsize{Node $j$}}
%\rput(-3.0,5.7) {\scriptsize{$T_{ji,1}$}}
%\rput(-2.3,5.8) {\scriptsize{$T_{ji,2}$}}
%\rput(-1.6,5.9) {\scriptsize{$T_{ji,3}$}}
%\rput(-0.2,6.15) {\scriptsize{$T_{ji,4}$}}
%\rput(2.5,6.55)  {\scriptsize{$T_{ji,K}$}}
%
%\rput(-3.9,4.7)   {\scriptsize{Node $i$}}
%\rput(-3.3,3.95)  {\scriptsize{$T_{ij,1}$}}
%\rput(-2.7,3.9)   {\scriptsize{$T_{ij,2}$}}
%\rput(-1.3,3.85)  {\scriptsize{$T_{ij,3}$}}
%\rput(-0.6,3.8)   {\scriptsize{$T_{ij,4}$}}
%\rput(3.2,3.7)   {\scriptsize{$T_{ij,K}$}}

%% copy
\includegraphics[scale=0.23]{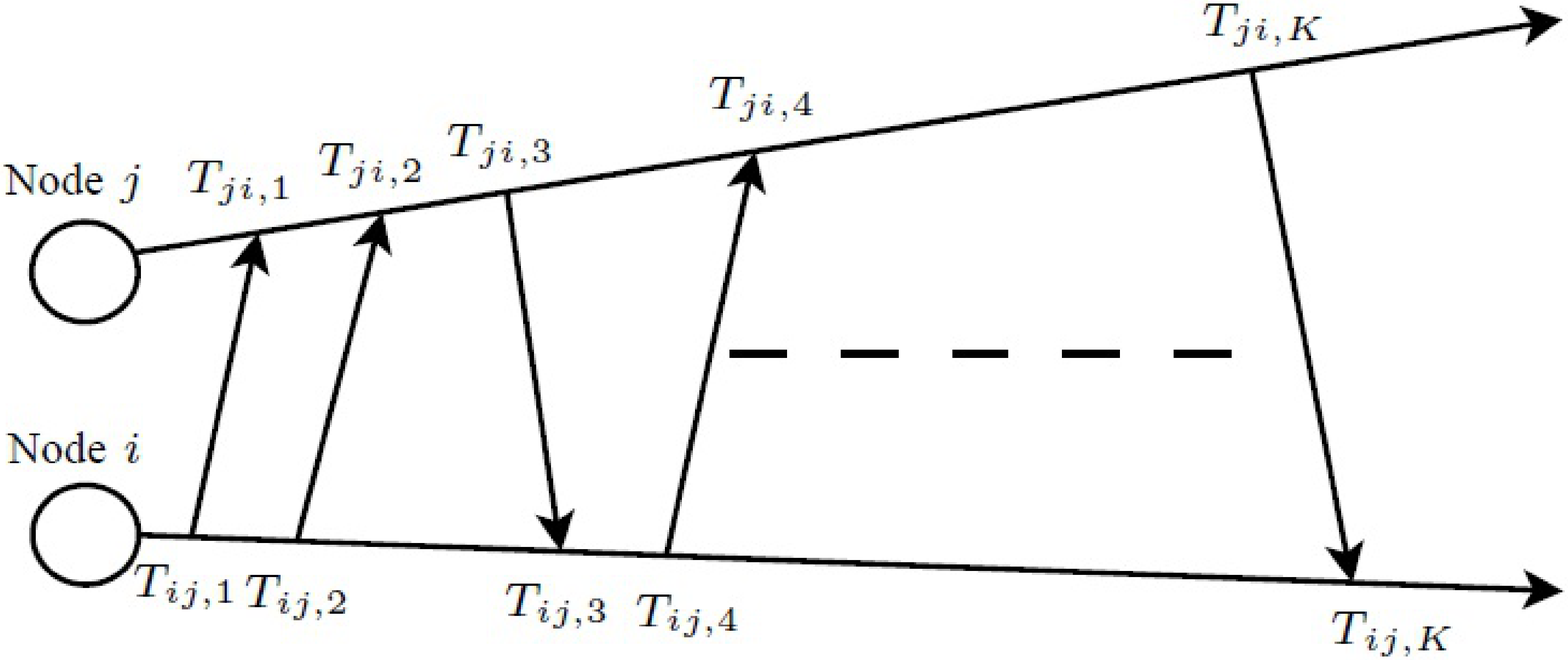}
\caption{A generalized Two-Way Ranging (TWR) between a pair of \emph{mobile} nodes, where the nodes transmit and receive, during which $K$ time stamps are recorded at the respective nodes. Similar to \cite{rajan2012icassp,rajan2013eusipco,rajanTSP1}, we levy no constraints on the sequence, direction or number of communications.}

\label{fig:figPairWise}
\end{figure}

\subsection{Data Model} We now consider a relaxed Two-Way Ranging (TWR) setup for collecting distance information as follows. Let a node pair $(i,j)$ within the network be capable of communicating with each other as shown in \figurename{\ref{fig:figPairWise}}. The nodes communicate $K$ messages back and forth, and the time of transmission and reception is registered independently at the respective nodes. The $k$th time stamp recorded at node $i$ when communicating with node $j$ is denoted by $T_{ij,k}$ and similarly at node $j$ the time stamp is $T_{ji,k}$. The direction of the communication is indicated by $E_{ij,k}$, where $E_{ij,k} = +1$ for transmission from node $i$ to node $j$ and $E_{ij,k} = -1$ for transmission from node $j$ to node $i$. Under ideal noiseless conditions, the propagation delay between the node pair at the $k$th time instant is $E_{ij,k}(T_{ij,k}- T_{ji,k})$,  and in conjunction with the polynomial approximation (\ref{eq:rangeDefinition2}), we have \begin{equation}\tau_{ij,k}= \ur_{ij} + \dur_{ij} T_{ij,k} + \ddur_{ij} T_{ij,k}^2 + \hdots = E_{ij,k}(T_{ji,k} - T_{ij,k}), \label{eq:rangeModelNoiseless}
\end{equation} where $\tau_{ij,k}\equiv\tau_{ij}(T_{ij,k})$ and without loss of generality we have replaced $t$ with $T_{ij,k}$.

\textit{{\bf Remark \arabic{remarkCounter}}: (Synchronized nodes): By replacing \emph{true} time $t$ by $T_{ij,k}$, we assume without loss of generality that $T_{ij,k}$ is in the neighborhood of $t_0=0$ and the propagation delay $\tau_{ij}$ is measured as a function of the local time at node $i$. Furthermore, we also assume that the clocks of these nodes are synchronized. This is a valid assumption since for an asynchronous network of mobile nodes, the clock parameters (up to first order) can be decoupled from the range parameters and estimated efficiently as shown in \cite{rajan2012icassp,rajan2013eusipco,rajanTSP1}.}
\stepcounter{remarkCounter}

In practice, the time measurements are also corrupted with noise and hence (\ref{eq:rangeModelNoiseless}) is \begin{eqnarray}
\ur_{ij}  + \dot{\ur}_{ij}(T_{ij,k}+ q_{i,k}) +\ddot{\ur}_{ij}(T_{ij,k}+ q_{i,k})^2 + \hdots\  \nonumber \\
        =  E_{ij,k}((T_{ji,k}+q_{j,k}) - (T_{ij,k}+q_{i,k}))
\label{eq:rangeModel_pairwise}
\end{eqnarray} where $q_{i,k}\sim \cN(0, \Sigma_i),\ q_{j,k} \sim \cN(0, \Sigma_j)$ are modelled as Gaussian \iid noise variables, plaguing the timing measurements at node $i$ and node $j$, respectively\footnote{Alternatively, the noise on the time markers can also be modeled as a uniformly random variable, typically rising from quantization errors. In addition, the proposed fixed variance model can be replaced a the distance-dependent variance model \cite{jia08}, which penalizes large inter-nodal distances.}. Rearranging the terms, we have \begin{equation}
\ur_{ij}  + \dot{\ur}_{ij}T_{ij,k} + \ddot{\ur}_{ij}T_{ij,k}^2 + \hdots =  E_{ij,k}(T_{ji,k} - T_{ij,k}) + q_{ij,k},
\end{equation} where \begin{equation}
q_{ij,k} = E_{ij,k}(q_{j,k} - q_{i,k}) - (2\ddur_{ij}T_{ij,k}q_{i,k} + \ddur_{ij}q^2_{i,k} + \hdots).
\end{equation} For wireless communication with $c= 3 \times 10^8$m/s, note that the modified range parameters are scaled by $c^{-1}$ (\ref{eq:rangeTranslation}). Furthermore, since the dynamic range model is proposed for a small time interval, the term $(2\ddur_{ij}T_{ij,k}q_{i,k} + \ddur_{ij}q^2_{i,k} + \hdots)$ is relatively small and subsequently the noise vector plaguing the measurements can be approximated as $q_{ij,k} \approx E_{ij,k}(q_{j,k} - q_{i,k})$ which begets \begin{equation} q_{ij,k} \sim\ \cN(0, \Sigma_{ij}), \label{eq:noisePairwise} \end{equation} where $\Sigma_{ij} = \Sigma_i + \Sigma_j$. Aggregating all $K$ packets, we have \begin{equation} \overbrace{\begin{bmatrix} \b1_K \quad \bt_{ij} \quad \bt^{\odot 2}_{ij} \quad \hdots\ \end{bmatrix}}^{\bA_{ij}} \overbrace{\begin{bmatrix} \ur_{ij} \\ \dot{\ur}_{ij} \\ \ddot{\ur}_{ij}  \\ \vdots\ \end{bmatrix}}^{\bthetau_{ij}} = \btau_{ij} + \bq_{ij}, \label{eq:pairwiseNormal}
\end{equation} where  \begin{eqnarray}
\btau_{ij}  &\triangleq& \be_{ij} \odot(\bt_{ji} - \bt_{ij}) \in \mathbb{R}^{K \times 1} \label{eq:tauDef},\\
\be_{ij}    &=& [E_{ij,1}, E_{ij,2}, \hdots, E_{ij,K}] \in \mathbb{R}^{K \times 1} \label{eq:eDef}, \\
\bt_{ij}    &=& [T_{ij,1}, T_{ij,2}, \hdots, T_{ij,K}] \in \mathbb{R}^{K \times 1} \label{eq:tDef}.
\end{eqnarray}

The known Vandermonde matrix $\bA_{ij} \in \mathbb{R}^{K \times L}$ contains the measured time stamps and is invertible if $T_{ij,k}$ is unique. The direction vector $\be_{ij}$ is encapsulated in the propagation delay $\btau_{ij}$ and $\bthetau_{ij} \in \mathbb{R}^{L \times 1}$ is a vector containing the unknown range parameters. The noise vector on this linear system is $\bq_{ij}=[q_{ij,1}, q_{ij,2}, \hdots\ q_{ij,K}]^T \in \mathbb{R}^{K \times 1}$, where $q_{ij,k}$ is given by (\ref{eq:noisePairwise}) and the corresponding covariance matrix is \begin{equation} \bSigma_{ij} \triangleq\ \mathbb{E}\left[ \bq_{ij}\bq_{ij}^T \right] = \Sigma_{ij}\bI_K\ \in \mathbb{R}^{K \times K}. \label{eq:noisePairwiseCov}\end{equation}
For a network of $N$ nodes, the normal equation (\ref{eq:pairwiseNormal}) can be extended to \begin{equation}
\overbrace{\begin{bmatrix} \bI_{\bar{N}} \otimes \b1_K \quad \bT \quad \bT^{\odot 2} \quad \hdots\ \end{bmatrix}}^{\bA} \overbrace{\begin{bmatrix} \bur \\ \bdur \\ \bddur  \\ \vdots\ \end{bmatrix}}^{\bthetau} = \btau + \bq, \label{eq:globalNormal} \end{equation} where \begin{eqnarray}
\bT     &=& \bdiag(\bt_{12}, \bt_{13}, \hdots \bt_{1N},\ \bt_{23}, \hdots ),  \in \mathbb{R}^{\bar{N}K \times \bar{N}} \\
\btau   &=& [\btau^T_{12}, \btau^T_{13}, \hdots \btau^T_{1N},\ \btau^T_{23}, \hdots]^T  \in \mathbb{R}^{\bar{N}K \times 1}
\end{eqnarray} contain the time stamp exchanges of the $\bar{N}$ \emph{unique} pairwise links in the network and $\bthetau \in \mathbb{R}^{\bar{N}L \times 1}$ contains the unknown range parameters for the entire network. The noise vector is $\bq=[\bq^T_{12}, \bq^T_{13}, \hdots, \bq^T_{1N},\ \bq^T_{23}, \hdots]^T \in \mathbb{R}^{\bar{N}K \times 1}$ and the covariance matrix is \begin{eqnarray}
\bSigma &\triangleq& \mathbb{E}\left[ \bq\bq^T \right] \in \mathbb{R}^{\bar{N}K \times \bar{N}K}.
\label{eq:noiseCov}\end{eqnarray}

\textit{{\bf Remark \arabic{remarkCounter}}: (Mobility of the nodes): In (\ref{eq:rangeModelNoiseless}), we implicity assumed that the nodes are relatively fixed during a time period of $\delta t_k = |T_{ij,k} - T_{ji,k}|$ \ie the propagation time of the message. This is a much weaker assumption compared to traditional TWR, where for a pair of fixed nodes (\ie\ $L=1$), the pairwise distance is assumed to be invariant for the total measurement period $\Delta T= |T_{ij,K}-T_{ij,1}|$. In reality, when the nodes are mobile, the distance at each $k$th time instant is dissimilar and this is inherently represented in the presented Dynamic ranging model.}
\stepcounter{remarkCounter}

\section{Dynamic ranging algorithm}  \label{sec:dynamicRangingAlgorithm} Suppose that we have collected all the TWR timing data in $\bA$ and $\btau$, then in this section we find an estimate for the unknown $\bthetau$ using the model (\ref{eq:globalNormal}). Given an estimate of $\bthetau$, the range coefficients $\btheta= [\br, \bdr, \bddr, \hdots]$ can be directly obtained from (\ref{eq:rangeTranslation}).

\subsection{Weighted Least Squares} Under the assumption that the covariance matrix $\bSigma$ is known, a Weighted Least Squares (WLS) solution $\hat{\bthetau}$ is obtained by minimizing the $l_2$ norm of the linear system (\ref{eq:globalNormal}), leading to \begin{eqnarray} \hat{\bthetau} &=& (\bA^T\bSigma^{-1}\bA)^{-1}\bA^T\bSigma^{-1}\btau \label{eq:globalLS} \end{eqnarray}  which is a valid solution if $K \ge L$ for each of the $\bar{N}$ pairwise links.  More generally, when  $L$ is unknown, an order recursive least squares \cite{Kay1993} can be employed to obtain the range coefficients for increasing values of $L$, until we reach an optimal polynomial fit for (\ref{eq:globalNormal}).

Furthermore, the \Cramer\ Rao lower Bound (CRB) \cite{Kay1993} for the least squares model ($\ref{eq:globalNormal}$) is \begin{equation}
\bSigma_{\thetau}=  (\bA^T\bSigma^{-1}\bA)^{-1}
\end{equation} and in combination with the range scaling (\ref{eq:rangeTranslation}), the CRB on $\btheta$ is given by \begin{equation} \bSigma_{\theta} \triangleq\  \bF(\bA^T\bSigma^{-1}\bA)^{-1}\bF  \label{eq:crb}
\end{equation} where \begin{equation}
\bSigma_{\theta}= \begin{bmatrix} \bSigma_r\ \qquad\ \qquad\ \qquad \\ \qquad\ \bSigma_{\dr}\ \qquad\ \qquad \\ \qquad\ \qquad\ \bSigma_{\ddr} \qquad \\ \qquad\ \qquad\ \qquad\ \ddots  \end{bmatrix} \label{eq:covTheta}
\end{equation} is the lowest variance attained by any unbiased estimate of the range parameters $\btheta= [\br^T, \bdr^T, \bddr^T, \hdots]^T$
and $\bF= \diag(\bff) \otimes \bI_{\bar{N}} \in \mathbb{R}^{\bar{N}L \times \bar{N}L}$. It is worth noting that (\ref{eq:globalLS}) achieves this lower bound. In addition, the lower bound is unaffected by the choice of direction vector $\be_{ij},\ \forall\ i,j \le N$, since all direction vectors are encapsulated in the measurement vector $\btau_{ij}$, which is not a part of the lower bound (\ref{eq:crb}).

\textit{{\bf Remark \arabic{remarkCounter}}: (Direction independence): In general, observe that the proposed solution (\ref{eq:globalLS}) is feasible for any direction marker $E_{ij,k}$, which is incorporated in $\btau$ (\ref{eq:tauDef}). Hence communication between the nodes could be arbitrary or one way, and need not be necessarily bi-directional. Note that, this is not true for an asynchronous network, where two-way communication is pivotal in jointly estimating the clock and range parameters \cite{rajanTSP1}. In addition, there is no pre-requisite on the number, sequence or direction of the communication links \cite{rajan2011camsap,rajan2012icassp,rajan2013eusipco,rajanTSP1}. Thus, the proposed solution is amenable to prevalent Two Way Ranging (TWR) protocols, such as classical pairwise communication \cite{ieee07}, passive listening and broadcasting \cite{wangTSP11}.}
\stepcounter{remarkCounter}

\subsection{Distributed Weighted Least Squares} If we consider independent pairwise communication between all the nodes, with no broadcasting, then the noise in each pairwise link is independent of each other and subsequently the covariance matrix (\ref{eq:noiseCov}) simplifies to \begin{eqnarray} \bSigma &=& \bdiag \left(\bSigma_{12}, \bSigma_{13}, \hdots \bSigma_{1N},\ \bSigma_{23}, \hdots \right). \label{eq:noiseCov_simple}\end{eqnarray} In which case, the centralized system (\ref{eq:globalNormal}) is a cascade of pairwise linear systems (\ref{eq:pairwiseNormal}) and subsequently (\ref{eq:globalLS}) is a generalized version of solving the distributed pairwise system for estimating the pairwise range parameters $\bthetau_{ij}$ \begin{eqnarray}
\hat{\bthetau}_{ij}  &=& \arg \min_{\bthetau_{ij}} \; \norm{\bSigma^{-1/2}_{ij}(\bA_{ij}\bthetau_{ij} - \btau_{ij})}^2  \nonumber \\
                    &=& (\bA_{ij}^T\bSigma_{ij}^{-1}\bA_{ij})^{-1}\bA_{ij}^T\bSigma^{-1}_{ij}\btau_{ij} \label{eq:pairwiseLS}
\end{eqnarray} which, similar to (\ref{eq:globalLS}), has a valid solution for $K \ge L$ for each pairwise link.

\section{Distances, Positions, Velocities\\ and Relative Kinematics}\label{sec:relativeKinematics}  In the previous section, we estimated $\bthetau$ which contains the solution to the unknown range derivatives $\btheta=\begin{bmatrix}\br, \bdr, \bddr, \hdots \end{bmatrix}$. Our next motive is to use these range derivatives to estimate the positions of the mobile nodes. When the nodes are in motion, similar to the pairwise range rates, the position vector of each node is also a Taylor series in time. However, exploiting piecewise linearity, we assume that the nodes are in linear motion with no acceleration, which is valid for a sufficiently small measurement period. (Note that despite this assumption, the pairwise distance is still non-linear.)

\subsection{Linear motion} Let the position of $N\ (N\ge P)$ nodes in a $P$-dimensional Euclidean space at the $k$th time instant be given by $\bX_k= [\bx_{1,k},\bx_{2,k}, \hdots\ \bx_{N,k}] \in \mathbb{R}^{P \times N}$, where $\bx_{i,k}\in \mathbb{R}^{P \times 1}$ is the position vector of the $i$th node at the $k$th message exchange. Furthermore, at time instant $t_0$, the $i$th node has velocity $\by_i \in \mathbb{R}^{P \times 1}$ and all such velocities are collected in $\bY= [\by_1,\by_2, \hdots\ \by_N] \in \mathbb{R}^{P \times N}$. Then, under a linear motion assumption, we have \begin{eqnarray}
\frac{\ensuremath{d} \by_i}{\ensuremath{d} t} = \bzero_P\ \quad  \forall\ i \le N. \label{eq:constantVelocity}
\end{eqnarray} Now, let $\Delta t_k = t_k- t_0$ where for the sake of notational convenience and without loss of generality, we assume $t_k= T_{ij,k}\ \forall\ k$, then the position matrix at the $k$th time instant is \begin{equation}
\bX_k =\  \bX + \Delta t_k\bY \label{eq:absoluteLinearModel} \end{equation}
where $\bX\triangleq \bX_0 = \begin{bmatrix}\bx_1,\bx_2, \hdots\, \bx_N \end{bmatrix}$ is the initial position matrix at time instant $t_0$ and $\bX_k$ only depends on the initial Position and Velocity (PV) of the nodes.

\subsection{Range derivatives} \label{sec:linearMotion} To estimate the position matrix $\bX_k$, we begin by stating explicit expressions for the range derivatives $\begin{bmatrix}\br, \bdr, \bddr, \hdots \end{bmatrix}$ in terms of $\bX,\bY$ under linear velocity assumption.

\begin{theorem} (Distance non-linearity) The pairwise distance $d_{ij}(t)$ between a node pair $(i,j)$ in $P\ge2$ dimensional Euclidean space is a non-linear function of time, \emph{even if the nodes are only in linear motion}. The range parameters $[r_{ij}, \dr_{ij}, \ddr_{ij}, \hdots\ ]$ at $t=t_0$ satisfy
\begin{subequations}\label{subeq:range} \begin{eqnarray}
r_{ij}    &=& \sqrt{\bx_i^T\bx_i + \bx_j^T\bx_j -2\bx_i^T\bx_j}, \label{eq:range} \\
\dr_{ij}  &=& r^{-1}_{ij}(\bx_i - \bx_j)^T(\by_i - \by_j),  \label{eq:drange}\  \\
\ddr_{ij} &=& r^{-1}_{ij}\left(\norm{(\by_i - \by_j)}^2 - \dr^2_{ij}\right). \label{eq:ddrange}
\end{eqnarray} \end{subequations}  \end{theorem} \begin{IEEEproof} See Appendix \ref{ap:rangeDerivation}. \end{IEEEproof}

Although these range parameters can be estimated up to the $(L-1)$th order efficiently (as demonstrated in Section \ref{sec:dynamicRangingAlgorithm}), in the rest of this article we utilize the information only up to $L=3$. Rearranging the equations for $r_{ij},\dr_{ij},\ddr_{ij}$, from $(\ref{subeq:range})$ we obtain \begin{subequations}\label{subeq:range1} \begin{eqnarray}
r_{ij}^2                   &=& (\bx_i - \bx_j)^T(\bx_i - \bx_j),  \label{eq:range1}\  \\
r_{ij}\dr_{ij}             &=& (\bx_i - \bx_j)^T(\by_i - \by_j),  \label{eq:drange1}\  \\
r_{ij}\ddr_{ij}+\dr^2_{ij} &=& (\by_i - \by_j)^T(\by_i - \by_j). \label{eq:ddrange1}\
\end{eqnarray} \end{subequations} Extending the above equations for all $N$ nodes, defining $\bg_{xx}=\diag(\bX^T\bX) \in \mathbb{R}^{N \times 1}, \bg_{xy}=\diag(\bX^T\bY) \in \mathbb{R}^{N \times 1}$ and $\bg_{yy}=\diag(\bY^T\bY) \in \mathbb{R}^{N \times 1}$, we have \begin{subequations}\label{subeq:range2} \begin{eqnarray}
\bR^{\odot 2}                           =& \bg_{xx}\b1^T_{N} + \b1_N\bg^T_{xx} -2\bX^T\bX,\;  \label{eq:range2}\  \\
\bR\odot\dot{\bR}                       =& \bg_{xy}\b1^T_N   + \b1_N \bg_{xy}^T - \bX^T\bY - \bY^T\bX,\;                         \label{eq:drange2}\  \\
\bR\odot\ddot{\bR}+\dot{\bR}^{\odot 2}  =& \bg_{yy}\b1^T_{N} + \b1_N\bg^T_{yy} -2\bY^T\bY,\;  \label{eq:ddrange2}\
\end{eqnarray} \end{subequations} where the square matrices $\bR = [r_{ij}]\in \mathbb{R_+}^{N \times N}$, $\dot{\bR} = [\dr_{ij}] \in \mathbb{R}^{N \times N}$ and $\ddot{\bR} = [\ddr_{ij}]\in \mathbb{R_+}^{N \times N}$ contain the initial pairwise ranges, range rates and rates of range rates, respectively. It is worth noting that $\bR$ and $\ddot{\bR}$ are Euclidean Distance Matrices (EDM)s, however, $\dot{\bR}$ although symmetric, may contain both positive and negative values and is thus not an EDM.

It is evident from (\ref{subeq:range2}) that without apriori knowledge of a few known PV, estimating the PVs of the network is an ill-posed problem and hence, we look to find solutions for the relative PV. Applying the centering matrix $\bP= \bI_N - N^{-1}\b1_N\b1_N^T \quad \in \mathbb{R}^{N \times N} $ on (\ref{subeq:range2}) and exploiting the property $\bP\b1_{N} = \bzero_N$, we have \begin{subequations} \label{subeq:range3} \begin{eqnarray}
\bB_{xx}  &=& \bP\bX^T\bX\bP,\;                        \label{eq:range3}\  \\
\bB_{xy}  &=& \bP(\bX^T\bY + \bY^T\bX)\bP,\;            \label{eq:drange3}\  \\
\bB_{yy}  &=& \bP\bY^T\bY\bP,  \label{eq:ddrange1}\;   \label{eq:ddrange3}\
\end{eqnarray} \end{subequations} where we for the sake of convenience, we have introduced \begin{subequations}\label{subeq:range4} \begin{eqnarray}
\bB_{xx} &\triangleq& -0.5\bP\bR^{\odot 2}\bP,\;                              \label{eq:range4}\  \\
\bB_{xy} &\triangleq& -\bP(\bR\odot\dot{\bR})\bP,\;                           \label{eq:drange4}\  \\
\bB_{yy} &\triangleq& -0.5\bP(\bR\odot\ddot{\bR}+ \dot{\bR}^{\odot 2})\bP.\;   \label{eq:ddrange4}\
\end{eqnarray} \end{subequations} The equations (\ref{eq:range3}) and (\ref{eq:ddrange3}) can now be used to estimate the initial relative positions and relative velocities of the nodes, via MDS. However, prior to applying MDS we first present definitions for the relative PVs.

\subsection{Relative framework} We define the relative PV vectors as an affine transformation of the corresponding absolute PV ($\bX_k, \bY$) \ie \begin{eqnarray} \bX_k &=& \bH_{x,k}\bXu_k + \bh_{x,k}\b1^T_N, \label{eq:absPos} \\
\bY &=& \bH_{y}\bYu + \bh_{y}\b1^T_N, \label{eq:absVel}
\end{eqnarray} where $\bXu_k$ is the relative position matrix of the nodes at $t_k$ up to a rotation $\bH_{x,k} \in \mathbb{R}^{P \times P}$ and translation $\bh_{x,k} \in \mathbb{R}^{P \times 1}$. Along similar lines, we define relative velocity as $\bH_y\bYu_k$ and relative velocity up to a rotation as $\bYu$, where $\bH_y \in \mathbb{R}^{P \times P}$ is an unknown rotation matrix. The relative velocity of the nodes $\bH_y\bYu$ is relative to the group velocity of the network, which is $\bh_y \in \mathbb{R}^{P \times 1}$. Under a linear velocity assumption (\ref{eq:constantVelocity}), the group velocity is the rate at which the relative translation vector varies with time \ie \begin{equation}
\bh_y = \Delta t_k^{-1} (\bh_{x,k}-\bh_{x,0}). \label{eq:groupVelocity}
\end{equation} Furthermore, the rotation matrices $\bH_{x,k},\bH_y$ are orthogonal \ie
\begin{equation} \bH^T_{x,k}\bH_{x,k}=\ \bH^T_y\bH_y=\ \bI_p \quad \forall\ 1 \le k \le K \label{eq:rotationMatrixProperty}.  \end{equation}  Now, substituting (\ref{eq:absPos}) and (\ref{eq:absVel}) in (\ref{eq:absoluteLinearModel}), and using the property (\ref{eq:groupVelocity}) we have \begin{eqnarray}
\bH_{x,k}\bXu_k &=&  \bH_{x,0}\bXu + \Delta t_k\bH_y\bYu, \label{eq:relativePositionk_}
\end{eqnarray} where for the sake of notational simplicity, we use $\bXu\triangleq\bXu_0$ to denote the relative position matrix at $t_0$.

Now observe that the translation vectors $\bh_{x,0}, \bh_{y}$ are unidentifiable from observations (\ref{subeq:range3}). Subsequently, we shall also see in the following section, that the solution to the relative PVs are independent of these translation vector and hence without loss of generality can be considered to be $\bzero_P$ for notational simplicity. Secondly, in order to have a meaningful interpretation of the relative position at the $k$th time instant (\ref{eq:relativePositionk_}), we must choose a reference coordinate system \eg $\bH_{x,0}=\bI$. To this end, without loss of generality and for notational simplicity, we have the following assumptions \begin{subequations} \label{eq:notionalAssumption}  \begin{eqnarray}
\bH_{x,0} &=& \bI_P, \label{subeq:Hx=I}\\
\bh_{x,0} &=& \bzero_P, \label{subeq:hx=0} \\
\bh_y &=&\bzero_P \label{subeq:hy=0}.
\end{eqnarray}\end{subequations} which simplifies (\ref{eq:relativePositionk_}) to \begin{equation}
\bX_k =\ \bXu + \Delta t_k\bH_y\bYu, \label{eq:relativePositionk}
\end{equation} where $\bX_k$ is the position of the nodes at the $k$th time instant up to a translation, under the assumption (\ref{eq:notionalAssumption}). More significantly, observe that the relative position at each $k$th time instant is only dependent on the relative PV and $\bH_{y}$. Hence in the following sections, our aim is to estimate $\bXu,\bYu$ and $\bH_y$, using the range parameters ($\bR, \dot{\bR}, \ddot{\bR}$) defined in (\ref{subeq:range4}) and estimated in Section \ref{sec:dynamicRangingAlgorithm}.

\subsection{Relative kinematic matrices} Substituting the expression for absolute PV from (\ref{eq:absPos}) and (\ref{eq:absVel}) respectively in (\ref{subeq:range3}), we have \begin{subequations}\label{subeq:range5} \begin{eqnarray}
\bB_{xx}  &=& \bP\bX^T\bX\bP = \bP\bX^T\bH_{x,0}^T\bH_{x,0}\bX\bP = \bXu^T\bXu,                 \label{eq:range5}\  \\
\bB_{xy}  &=& \bP(\bX^T\bY + \bY^T\bX)\bP  \nonumber \\
          &=& \bP(\bXu^T\bH^T_{x,0}\bH_y\bYu + \bYu^T\bH^T_y\bH_{x,0}\bXu)\bP \nonumber \\
          &=& \bXu^T\bH_{y}\bYu + \bYu^T\bH^T_{y}\bXu,\;                              \label{eq:drange5}\  \\
\bB_{yy}  &=& \bP\bY^T\bY\bP = \bP\bY^T\bH_y^T\bH_y\bY\bP  = \bYu^T\bYu,              \label{eq:ddrange5}\
\end{eqnarray} \end{subequations} where we use the property (\ref{eq:rotationMatrixProperty}) in (\ref{eq:range5}) and (\ref{eq:ddrange5}), and the assumption (\ref{subeq:Hx=I}) in (\ref{eq:drange5}). $\bB_{xx}$ and $\bB_{yy}$ are Gramian matrices of the relative PVs and the expression for $\bB_{xy}$ is the Lyapunov-like linear matrix equation \cite{braden1998}. It is worth noting that the relative kinematic equations $\bB_{xx}, \bB_{xy}, \bB_{yy}$ are dependent only on the relative PVs and the unique rotation matrix at time $t_0$. For an alternative derivation of the relative kinematic matrices, refer to Appendix \ref{ap:Bderivation}.

Given an estimate of the range matrices, \ie $\widehat{\bR}, \widehat{\dot{\bR}}, \widehat{\ddot{\bR}}$, either using (\ref{eq:globalLS}) or alternative methods, an estimate of the relative kinematic matrices, \ie  $\widehat{\bB}_{xx}, \widehat{\bB}_{xy}, \widehat{\bB}_{yy}$ can be readily obtained using (\ref{subeq:range4}). Following which, we aim to estimate the relative position using (\ref{eq:range5}), the relative velocity using (\ref{eq:ddrange5}) and the unknown velocity rotation matrix $\bH_y$ using (\ref{eq:drange5}).

\section{Algorithms} \label{sec:algorithms}

\subsection{Relative positions ($\bXu$) and Relative velocities ($\bYu$)} \label{sec:relativePV}  An estimate of the relative PV can be directly obtained by the spectral decomposition of the matrices $\bB_{xx}, \bB_{yy}$. Let
\begin{eqnarray}
\widehat{\bB}_{xx} &=& \bU_x\bLambda_x\bU_x^T, \\
\widehat{\bB}_{yy} &=& \bU_y\bLambda_y\bU_y^T,
\end{eqnarray} where $\bU_x,\bU_y \in \mathbb{R}^{N \times N}$ contain the eigenvectors and the diagonal matrices $\bLambda_x, \bLambda_y \in \mathbb{R}^{N \times N}$ contain the increasingly ordered eigenvalues of the matrices $\widehat{\bB}_{xx}, \widehat{\bB}_{yy}$ respectively. Then, for a $P$-dimensional setup, an estimate of the relative positions $\bXu$ and relative velocities $\bYu$ of the nodes up to a rotation is then \begin{eqnarray}
\widehat{\bXu} &=& \underline{\bLambda}^{1/2}_x\underline{\bU}^T_x, \label{eq:relativePosition} \\
\widehat{\bYu} &=& \underline{\bLambda}_y^{1/2}\underline{\bU}_y^T, \label{eq:relativeVelocity}
\end{eqnarray} where $\underline{\bLambda}_x,\underline{\bLambda}_y  \in \mathbb{R}^{P \times P}$ contain the first $P$ nonzero eigenvalues and $\underline{\bU}_x, \underline{\bU}_y \in \mathbb{R}^{N \times P}$ contain the corresponding eigenvectors.

Relative positioning (\ref{eq:relativePosition}) from pairwise distance measurements using MDS is a well known technique \cite{borg97}. However, our contribution is the definition and estimation of relative velocities , \ie (\ref{eq:absVel}) and (\ref{eq:relativeVelocity}) respectively.

\subsection{Rotation matrix $\bH_y$} \label{sec:relativeRotation} The estimate of the relative velocity $\bYu$ up to an arbitrary rotation gives no information on the direction of the nodes in an anchorless scenario. Hence, it is important to estimate the relative velocities \wrt the orientation of the initial positions \ie $\bH_y$. Substituting the estimates of $\bB_{xy}, \bXu, \bYu$ from (\ref{eq:drange4}), (\ref{eq:relativePosition}) and (\ref{eq:relativeVelocity}) respectively in (\ref{eq:drange5}), we have \begin{eqnarray}
\widehat{\bB}_{xy} &=& \widehat{\bXu}^T\bH_{y}\widehat{\bYu} + \widehat{\bYu}^T\bH^T_{y}\widehat{\bXu},
\label{eq:rotationModel} \end{eqnarray} where $\bH_y$ is the unknown unitary matrix which can be estimated by minimizing the cost function \begin{eqnarray}
\widehat{\bH}_y= \arg \min_{\bH_y} \Big( \norm{\widehat{\bB}_{xy}- (\widehat{\bXu}^T\bH_y\widehat{\bYu} + \widehat{\bYu}^T\bH^T_y\widehat{\bXu})}^2 \Big), \label{eq:rotationCostFunction} \end{eqnarray} where $\widehat{\bH}_y$ is an estimate of $\bH_y$. Now, vectorizing (\ref{eq:rotationModel}) and rearranging the terms, we have \begin{eqnarray}
\bb_{xy} &=& (\widehat{\bYu}^T \otimes \widehat{\bXu}^T)\ \text{vec}(\bH_y) +(\widehat{\bXu}^T \otimes \widehat{\bYu}^T)\ \text{vec}(\bH^T_y) \nonumber\\ &=& (\bI_{N^2} + \bJ)(\widehat{\bYu}^T \otimes \widehat{\bXu}^T)\ \text{vec}(\bH_y)   \nonumber \\
&=& \bG\text{vec}(\bH_y),
\end{eqnarray} where  $\bb_{xy}= \text{vec}(\widehat{\bB}_{xy})$ is a vector of the known measurement matrix $\widehat{\bB}_{xy}$ from (\ref{eq:drange3}) and $\bJ \in \mathbb{R}^{N^2 \times N^2}$ is an orthogonal permutation matrix such that $\bJ\text{vec}(\bH_y) = \text{vec}(\bH_y^T) $. The unknown unitary matrix $\bH_{xy}$ can then be obtained by reformulating (\ref{eq:rotationCostFunction}) and solving \begin{eqnarray}
\widehat{\bH}_y =\ \argmin_{\bH_y}\ \norm{\bG\text{vec}(\bH_y)- \bb_{xy}} ^2=\ (\bG^T\bG)^{-1}\bG\bb_{xy}
\label{eq:vecHy}, \end{eqnarray} which has a feasible solution for $N \ge P$. The proposed solution does not exploit the orthogonailty property of the unknown rotation matrix $\bH_y$. Hence, more optimal solutions are feasible  \cite{viklands2006} by solving the constrained cost function \begin{equation}
\widehat{\bH}_y = \argmin_{\bH_y}\ \norm{\bG\text{vec}(\bH_y)- \bb_{xy}} ^2\;\;\text{s.t}\quad \bH_y^T\bH_y= \bI_P.
\end{equation}

\section{Relative position at time instant $k$} We now briefly summarize the steps to find the relative position at discrete time instances using the time stamp measurements discussed in Section \ref{sec:dynamicRanging}.

\subsection{Dynamic MDS} Given the noisy time stamps $\widehat{T}_{ij,k}= T_{ij,k} + q_{i,k}, \forall\ (i,j)$ node pairs in the network and $\forall\ 1\le k\le K$ time instances, the relative position of the nodes at the $k$th time instance can be estimated as follows. \begin{itemize}
\item Solve for an estimate of the Range derivatives $\widehat{\bR}, \widehat{\dot{\bR}}, \widehat{\ddot{\bR}}$ using Dynamic ranging (\ref{eq:globalLS}).
\item Using these estimated range derivatives, construct the relative kinematic matrices $\widehat{\bB}_{xx}, \widehat{\bB}_{xy}, \widehat{\bB}_{yy}$ defined in (\ref{eq:range4}).
\item Obtain an estimate of the relative PV and unitary matrix from (\ref{eq:absPos}), (\ref{eq:absVel}) and (\ref{eq:vecHy}) respectively. Then, using (\ref{eq:relativePositionk}) and defining $\Delta \hat{t}_k= \widehat{T}_{ij,k}- \widehat{T}_{ij,0}$, the relative position at the $k$th time instant is \begin{equation}
    \widehat{\bX}_{k,dr} =\ \widehat{\bXu} + \Delta \hat{t}_k \widehat{\bH}_y\widehat{\bYu}.\end{equation}
\end{itemize}

\subsection{Classical MDS} Alternatively, the relative positions of the nodes can also be estimated using Classical MDS (CMDS). Let $\bD_k \triangleq\ c[\tau_{ij,k}] \in \mathbb{R}^{N \times N}$ be the EDM at each discrete time instant $k$ where $\tau_{ij}= T_{ij,k} - T_{ji,k}$ and $\widehat{\bD}_k \triangleq\ c[\tau_{ij,k}+ q_{ij,k}] $ be the corresponding noisy estimate where $q_{ij,k}$ is the noise plaguing the measurements as shown in (\ref{eq:noisePairwise}). Let $-0.5\bP(\widehat{\bD}^{\odot 2}_k)\bP = \bar{\bU}_k\bar{\bLambda}_k\bar{\bU}_k^T$ be an eigenvalue decomposition, then the solution to the relative position is \begin{equation}
\widehat{\bXu}_{k,cmds}= \bar{\underline{\Lambda}}_k^{1/2}\bar{\underline{\bU}}_k^T
\end{equation} where $\bar{\underline{\Lambda}}_k \in \mathbb{R}^{P\times P}$ contain the first $P$ nonzero eigenvalues and $\bar{\underline{\bU}}_k \in \mathbb{R}^{N\times P}$ the corresponding eigenvectors.

Note that the relative position estimate using CMDS \ie $\widehat{\bXu}_{k,cmds}$ is up to an arbitrary rotation and translation, where as $\widehat{\bX}_{k,dr}$ yields the relative position of the nodes up to a translation alone.

\section{Simulations} \label{sec:simulations} Simulations are conducted to evaluate the performance of the proposed solutions. We consider a cluster of $N=10$ nodes in $P=2$ dimensions, whose coordinates $\bX$ and velocities $\bY$ are arbitrarily chosen as
\begin{eqnarray}
\bX &=&
  \begin{bmatrix}
   -382&   735&   959&   630&   800\\
   9&     7&   727&   366&  -858
  \end{bmatrix}, \nonumber \\
\bY &=&
  \begin{bmatrix}
     -6&     8&    -1&   -10&     3 \\
     8&    -9&    -7&    -2&    -8
  \end{bmatrix}. \nonumber
\label{eq:xyValues}
\end{eqnarray} Without loss of generality, we assume that all nodes employ one-way communication, \ie $\be_{ij} = \b1_K, \forall\ i,j \le N$. Furthermore, all nodes communicate with each other within the same time interval $\Delta T = [T_{ij,1}, T_{ij,K}] = [-3, 3]$ seconds and the transmit time markers are chosen to be linearly spaced within this interval. We consider a classical pairwise communication scenario, where all the pairwise communications are independent of each other and thus $\bSigma=\sigma^2\bI_{\bar{N}K}$.

The metric used to evaluate the performance of the range parameters is the Root Mean Square Error (RMSE), given by RMSE$(\bz)= \sqrt{N^{-1}_{exp} \sum^{N_{exp}}_{n=1}\norm{\hat{\bz}(n)-\bz}^2}$, where $\hat{\bz}(n)$ is the $n$th estimate of the unknown vector $\bz \in \mathbb{R}^{\bar{N} \times 1}$ during $N_{exp}=1000$ Monte Carlo runs. To qualify these estimates, the square Root of the \Cramer\ Rao Bound (RCRB) is plotted along with the respective RMSE. We also use the same metric for evaluating the rotation $\bh_{xy}=\text{vec}(\bH_{xy})$.

However, since the relative PVs (\bXu, \bYu) and $\bX_k$ are known only up to an arbitrary rotation, we define the RMSE for these matrices as RMSE$(\bZ)= \sqrt{N^{-1}_{exp} \sum^{N_{exp}}_{n=1}\norm{\text{vec}(\bH\widehat{\bZ}(n)-\bZ\bP)}^2}$, where $\bP$ is the centering matrix and $\bH$ is the optimal Procrustes rotation, given the matrix $\bZ$ and the corresponding estimate $\widehat{\bZ}(n)$ of the $n$th Monte Carlo run. See Appendix \ref{ap:procrustes}. For the relative PV the \Cramer\ Rao bounds are derived (Appendix \ref{ap:crbPV}) and the corresponding RCRBs are plotted along with the RMSEs.

\subsection{Varying Number of communications ($K$)} 
\begin{figure}
  \centering
  \begin{subfigure}[b]{0.5\textwidth}
    \includegraphics[scale=0.6]{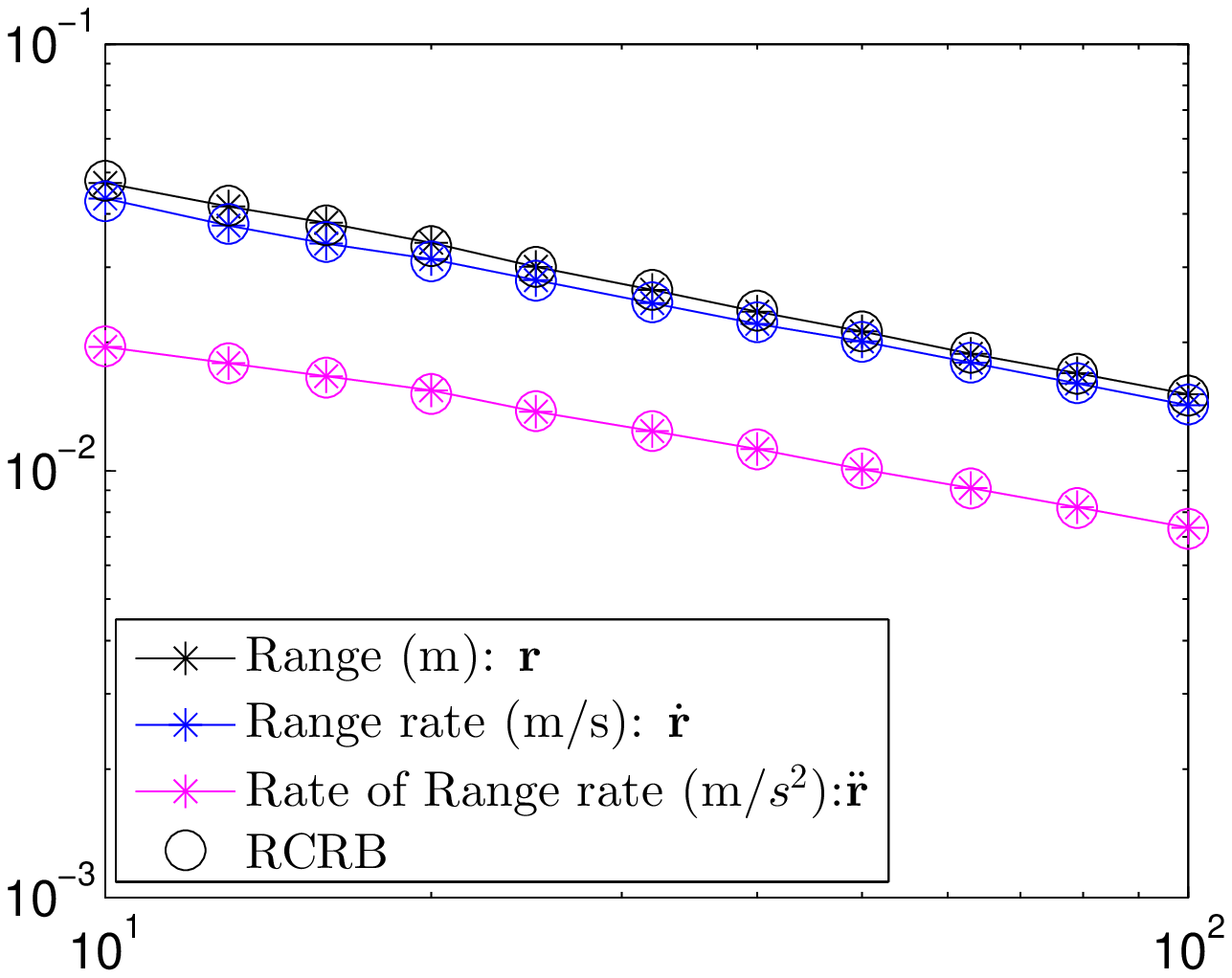}
    \rput(4.6,0.5){\small{Number of two-way communications (K)}}
    \rput*{90}(0, 3.6){\small{RMSE of range coefficients}}
    \caption{} \label{fig:ranges}
  \end{subfigure} \\
  \begin{subfigure}[b]{0.5\textwidth}
    \includegraphics[scale=0.6]{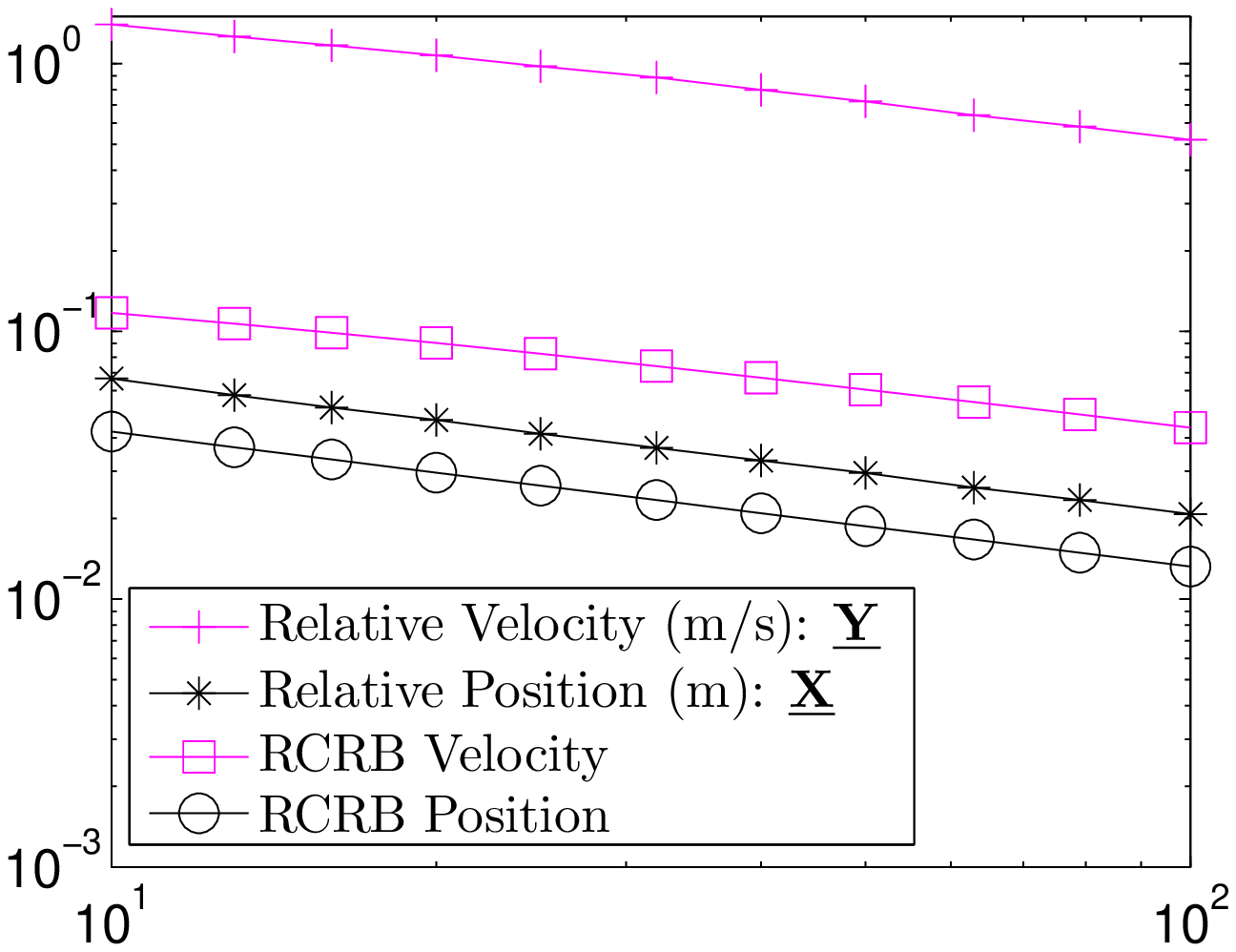}
    \rput(4.6,0.5){\small{Number of two-way communications (K)}}
    \rput*{90}(0, 3.6){\small{RMSE of relative PVs}}
    \caption{} \label{fig:pv}
  \end{subfigure} \\
  \begin{subfigure}[b]{0.5\textwidth}
    \includegraphics[scale=0.6]{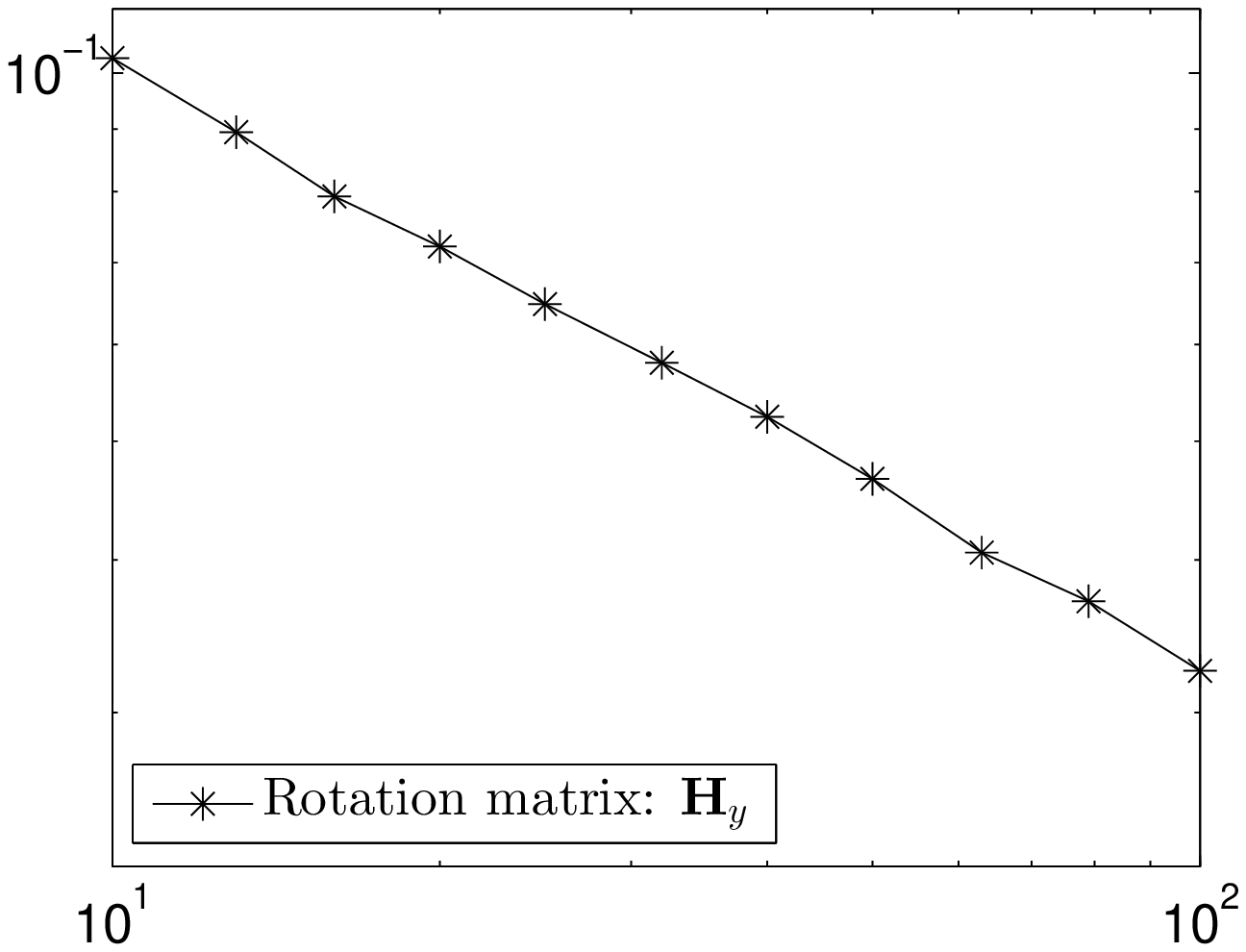}
    \rput(4.6,0.5){\small{Number of two-way communications (K)}}
    \rput*{90}(0, 3.6){\small{RMSE of velocity rotation}}
    \caption{} \label{fig:hxy}
  \end{subfigure}
  \caption{\small RMSEs of (a) range parameters, (b) relative position, relative velocity and (c) relative rotation matrix for varying number of communications ($K$) between the nodes for $\sigma=0.1$ meters}
\end{figure}

The dynamic ranging algorithm  (\ref{eq:globalLS}) is implemented for $L=4$, where the number of communications $K$ is varied from $10$ to $100$. The noise on the propagation delays is $\sigma=0.1$ meters, which is typical in classical TWR \cite{patwari2003} or in conventional anchored MDS-based velocity estimation using Doppler measurements \cite{wei10}. \figurename\ \ref{fig:ranges} shows the RMSE of the first $3$ range coefficients (which are relevant for estimating the relative velocities) achieving the RCRB asymptotically. The PV estimates are obtained using these range coefficients via (\ref{eq:relativePosition}), (\ref{eq:relativeVelocity}) and the corresponding RMSEs are plotted in \figurename\ \ref{fig:pv}, along with respective RCRBs. Furthermore, the RMSEs of the relative rotation matrix $\bH_{xy}$ estimate (\ref{eq:vecHy}) is shown in \figurename\ \ref{fig:hxy}, where the relative position and velocity estimates are used.

\subsection{Varying noise on time measurements ($\sigma$)} 
\begin{figure}
  \centering
  \begin{subfigure}{0.5\textwidth}
    \includegraphics[scale=0.6]{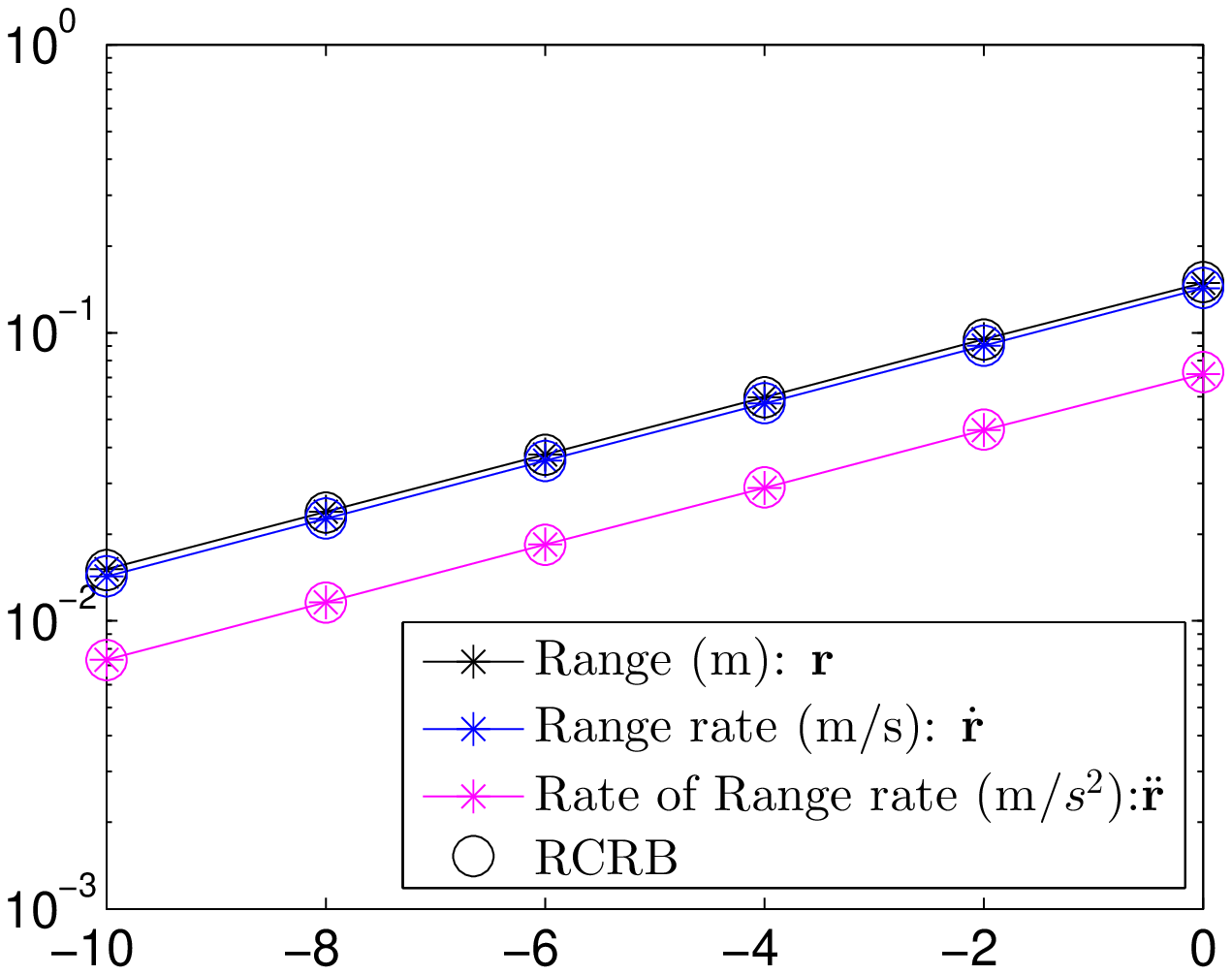}
    \rput(4.6,0.5){\small{$10\log_{10}(\sigma)$ [dB meter]}}
    \rput*{90}(0, 3.6){\small{RMSE of range coefficients}}
    \caption{} \label{fig:ranges_snr}
  \end{subfigure} \\
  \begin{subfigure}{0.5\textwidth}
    \includegraphics[scale=0.6]{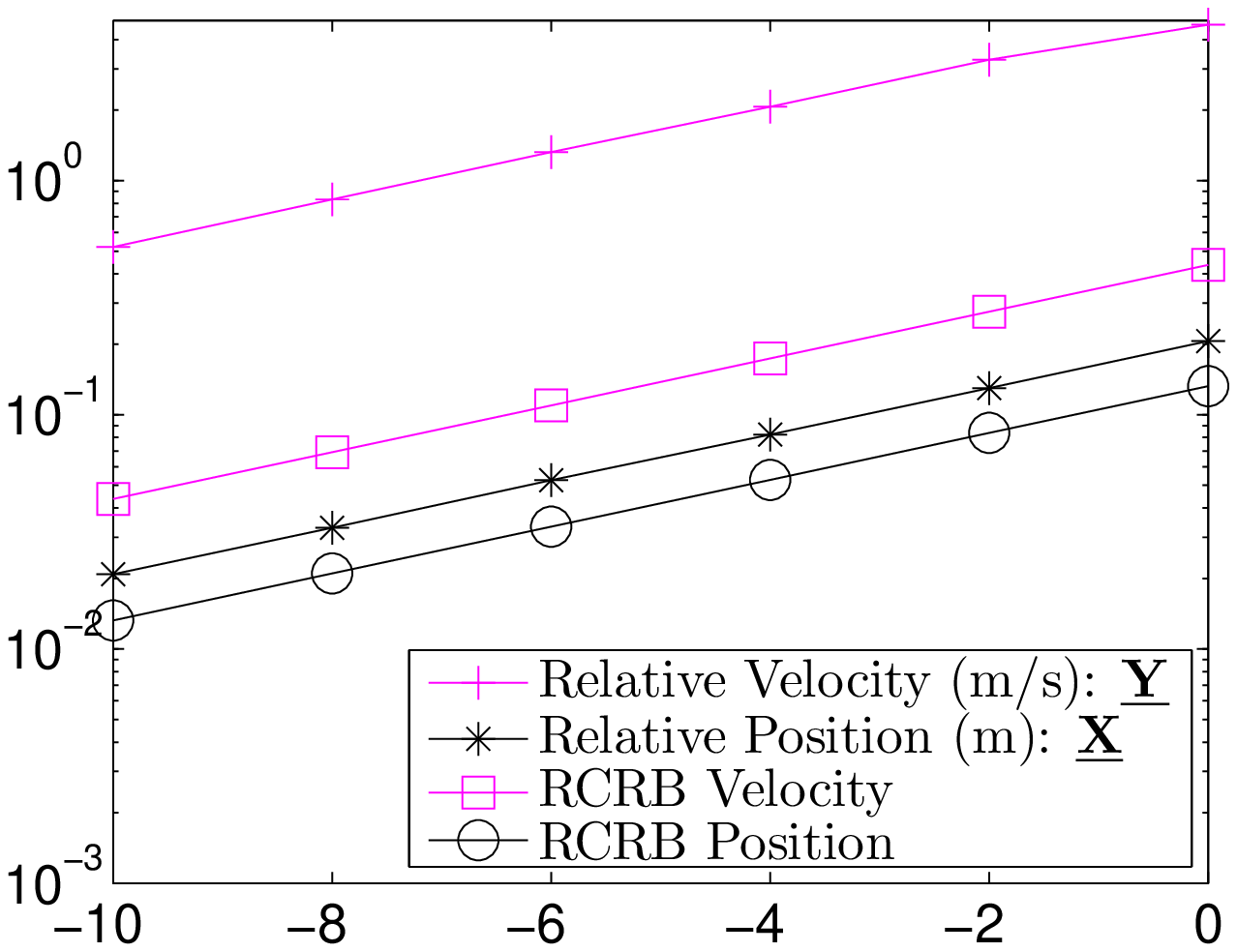}
    \rput(4.6,0.5){\small{$10\log_{10}(\sigma)$ [dB meter]}}
    \rput*{90}(0, 3.6){\small{RMSE of relative PVs}}
    \caption{} \label{fig:pv_snr}
  \end{subfigure}\\
  \begin{subfigure}{0.5\textwidth}
    \includegraphics[scale=0.6]{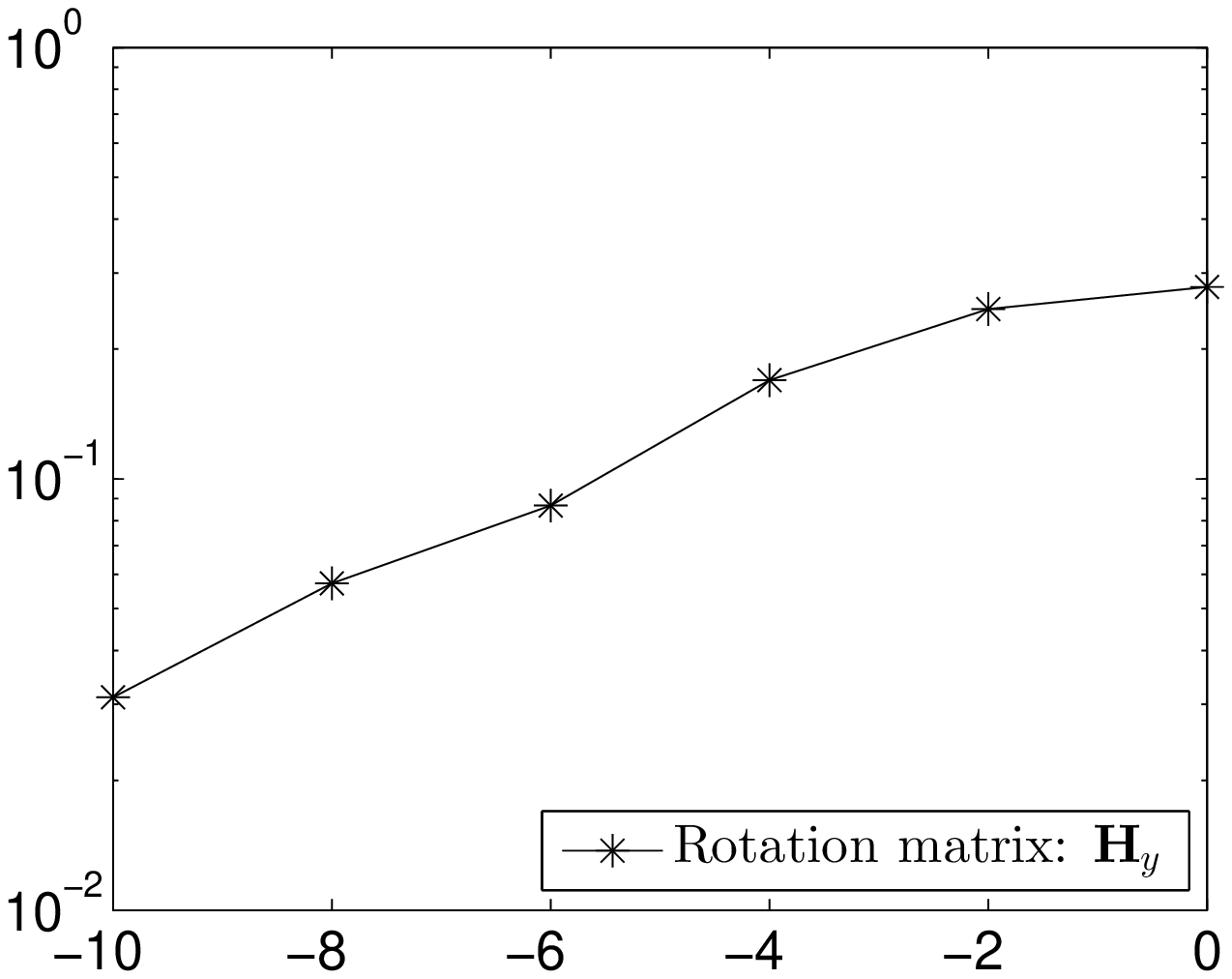}
    \rput(4.6,0.5){\small{$10\log_{10}(\sigma)$ [dB meter]}}
    \rput*{90}(0, 3.6){\small{RMSE of velocity rotation}}
    \caption{} \label{fig:hxy_snr}
  \end{subfigure}
  \caption{\small RMSEs (a) range parameters, (b) relative position, relative velocity and (c) relative rotation matrix for varying noise ($\sigma$) on the Time measurements with number of communication $K=100$ }
\end{figure}

A second experiment is carried out by varying $\sigma$ in the range $[-10, 0]$ dB meters for a fixed number of communications $K=100$. The RMSEs of the range coefficients obtained via the dynamic ranging algorithm (\ref{eq:globalLS}) are plotted in \figurename\ \ref{fig:ranges_snr}, which achieve the RCRB asymptotically. The RMSEs on the relative PV are shown in \figurename\ \ref{fig:pv_snr}, and the RMSE of the relative rotation matrix is presented in \figurename\ \ref{fig:hxy_snr}, in addition to the corresponding RCRBs. To the best of the our knowledge, given the novelty of the data model and the corresponding solutions, there are no other relative velocity estimators available for comparison.

%Note that, the rotation matrix estimate performs stably for high SNRs and is limited by the variance of the relative velocity estimate (\bYu) for lower SNRs.

\begin{figure}
  \centering
  \includegraphics[scale=0.6]{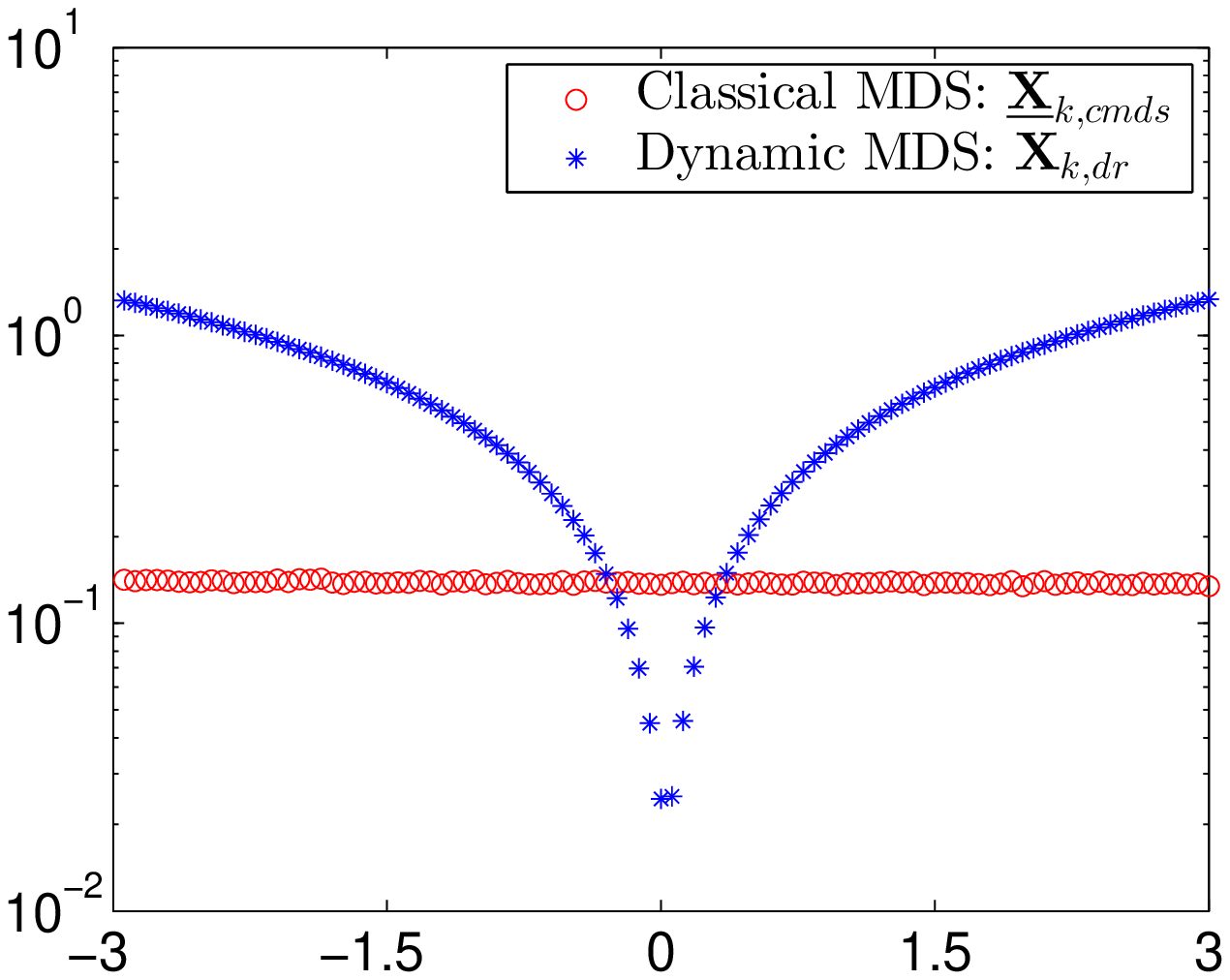}
  \rput(-4,-0.15){Time [seconds]}
  \rput*{90}(-8.8, 3.6){\small{RMSE of relative positions}}
  \caption{\small{RMSE of relative positions at discrete time instances $t_k$ during the time interval $\Delta T= [-3,3]$ with $K=100$ for $\sigma= 0.1$ meters}}\label{fig:Xrk}
\end{figure}

\subsection{Relative position error over time} Figure \ref{fig:Xrk} shows the RMS plots for $\bXu_{k,\text{cmds}}$ and $\bX_{k,\text{dr}}$ for a time duration $\Delta T= [-3,3]$ with Gaussian noise of $\sigma=0.1$ meters on the distance measurements. The $\bXu_{k,\text{cmds}}$ estimate steadily achieves a constant RMSE, which is expected since CMDS is independently applied at each $k$th time instant, to estimate the relative positions of the nodes. On the contrary, the relative position estimation via dynamic ranging betters this estimate around $t_0$, where the improvement of up to a factor $\sqrt{K}$ is primarily due to averaging over $K$ measurements. However, the error estimate of $\bX_{k,\text{dr}}$ increases as we move away from $t_0$, which is typical of Taylor series approximation. In addition, the poor performance of the Classical MDS based algorithm for relative velocity estimate (see \figurename\ (\ref{fig:pv}), \figurename\ (\ref{fig:pv_snr})) also hampers the solution for $\bX_{k,dr}$. An improved estimate for relative velocity estimation is feasible, which will be addressed in future work.

%It is evident that for both varying number of communications and varying SNR, the variance of the relative velocity up to a rotation ($\bYu$) is dominant by orders of magnitude compared to the relative rotation estimate. Hence, the RMSEs of the relative velocities $\bH_{xy}\bYu$ \wrt the node position is not shown.

\section{Conclusions} A novel framework is proposed to estimate the relative positions up to a rotation for an \emph{anchorless network of mobile nodes} without the use of Doppler measurements. The proposed least squares based dynamic ranging algorithm employs a classical Taylor series based approximation, which extracts pairwise distance derivatives at a given time instant efficiently. Under a linear velocity assumption, we show that the time-varying relative positions can be estimated from the derivatives of the pairwise distances. The initial relative positions, relative velocities and a unique rotation matrix are sufficient to describe the relative motion of the nodes during a small time interval. Subsequently, closed form MDS-based solutions are presented to jointly estimate the relative positions and relative velocities of the nodes. In addition, the unique rotation matrix which relates the direction of the relative motion \wrt the relative position is also estimated via least squares. The \Cramer\ Rao bounds are also derived for the range parameters, and the relative PV and simulations are conducted to verify and analyze the performance of the proposed least squares estimators. The presented solutions are suited for autonomous networks with minimal a priori knowledge, where the positions and velocities need to be estimated at \emph{cold start}. In practice, over longer durations, the estimated parameters can be readily extended to both relative and absolute tracking, which is beyond the scope of this article and will be addressed in a follow-up work.

\appendices
\section{Distance Non-linearity} \label{ap:rangeDerivation} Consider an arbitrary pair of mobile nodes with time-varying positions $\{\bar{\bx}_i(t), \bar{\bx}_j(t)\}$ and constant velocities $\{\by_i, \by_j\}$. In addition, we define the position of the nodes at $t=t_0$ as  $\{\bx_i, \bx_j\}$. To show that the time-varying $d_{ij}(t)$ is an infinitely differentiable function we derive the first few derivatives of $d_{ij}(t)$ \wrt time. By definition, the initial pairwise distance between the nodes is the Euclidean norm \begin{eqnarray}
r_{ij}\triangleq\ d_{ij}(t_0)  &=& \norm{\bx_{i} - \bx_{j}}
\label{eq:rangeL0}
\end{eqnarray}  \subsubsection{First order $\dr_{ij}$} From (\ref{eq:rangeL0}), we can compute the first-order range parameter as   \begin{eqnarray} \dr_{ij}  &=& \frac{\ensuremath{d}\ }{\ensuremath{d} t}d_{ij}(t)  \nonumber\\
              &=& \frac{1}{2r_{ij}} \frac{\ensuremath{d} }{\ensuremath{d} t} \Big((\bar{\bx}_i(t) - \bar{\bx}_j(t))^T(\bar{\bx}_i(t) - \bar{\bx}_j(t)) \Big) \nonumber \\
              &=& \frac{1}{r_{ij}} \Big(\by_i^T\bx_i + \by_j^T\bx_j - \by_i^T\bx_j - \by_j^T\bx_i \Big) \nonumber \\
              &=& r^{-1}_{ij} (\by_i- \by_j)^T(\bx_i - \bx_j)
\label{eq:rangeL1}
\end{eqnarray}
\subsubsection{Second order $\ddr_{ij}$}
Similarly, under the assumption of constant velocities, the second-order range parameter using (\ref{eq:rangeL0}) is \begin{eqnarray}
\ddr_{ij} &=& \frac{\ensuremath{d^2}}{\ensuremath{d} t^2}d_{ij}(t)  \nonumber \\
              &=& -r_{ij}^{-2}\dr_{ij} \Big((\by_i- \by_j)^T(\bx_i - \bx_j)\Big) \nonumber \\
                  & &  + r_{ij}^{-1} \frac{\ensuremath{d} }{\ensuremath{d} t}\Big((\by_i- \by_j)^T(\bar{\bx}_i(t) - \bar{\bx}_j(t))\Big)  \nonumber \\
              &=& -r_{ij}^{-1}\dr^2_{ij} + r_{ij}^{-1} (\by_i- \by_j)^T(\by_i - \by_j)  \nonumber \\
              &=& r^{-1}_{ij} \Big(\norm{\by_i - \by_j}^2-  \dr^2_{ij}\Big)
\label{eq:rangeL2}
\end{eqnarray}

\subsubsection{Third order $\dddr_{ij}$}
The third-order derivative of the range parameter under linear motion (\ref{eq:rangeL0}) yields   \begin{eqnarray}
\dddr_{ij} &=& \frac{\ensuremath{d^3}}{\ensuremath{d} t^3}d_{ij}(t) \nonumber \\
              &=& -r_{ij}^{-2}\dr_{ij} (\norm{\by_i- \by_j}^2 - \dr^2_{ij})- r^{-1}_{ij}\frac{\ensuremath{d}^2 }{\ensuremath{d} t^2}(d^2_{ij}(t))\nonumber \\
              &=& -r_{ij}^{-1}\dr_{ij}\ddr_{ij} - 2r^{-1}_{ij}\dr_{ij}\ddr_{ij}  \nonumber \\
              &=& -3r^{-1}_{ij}\dr_{ij}\ddr_{ij}
\label{eq:rangeL3}
\end{eqnarray} The higher-order range derivatives can be derived along similar lines.

\section{Alternative derivation for $\bB_{xx},\bB_{xy}\, \bB_{yy} $}  \label{ap:Bderivation} With an abuse of notation, let $\bD(t) \in \mathbb{R}^{N \times N}$ be the time-varying Euclidean Distance Matrix (EDM) for a network of $N$ nodes in $P$-dimensional Euclidean space and let \begin{equation} \bB(t) = -0.5\bP\bD(t)^{\odot 2}\bP, \end{equation} where $\bP= \bI_N - N^{-1}\b1_N\b1^T_N$ is the centering matrix. Then observe that at $t=t_0$, \begin{equation} \bB(t_0) \triangleq\ \bB_{xx} = \bXu^T\bXu\end{equation} and the subsequent first derivative is
\begin{eqnarray} \bB_{xy} &\triangleq& \frac{\ensuremath{d} \bB(t)}{\ensuremath{dt}}\ \triangleq\ -\bP \Big( \bD(t) \odot \dot{\bD}(t)\Big)\bP\Big|_{t=t_0} \nonumber \\ &=& \bXu^T\bH_{xy}\bYu + \bYu^T\bH^T_{xy}\bXu. \end{eqnarray} A step further, differentiating again \wrt time and substituting $t=t_0$ we have \begin{equation} \frac{\ensuremath{d}^2 \bB(t)}{\ensuremath{dt^2}}\Big|_{t=t_0} \triangleq\ \bB_{yy}\triangleq\ -0.5\bP(\bR\odot\ddot{\bR} + \dot{\bR}^{\odot 2})\bP = \bYu^T\bYu \label{eq:dbyy} \end{equation} where $\dot{\bR}=[\dot{r}_{ij}] \in \mathbb{R}^{N \times N}$ and $\ddot{\bR} =[\ddot{r}_{ij}]\in \mathbb{R}_+^{N \times N}$ which, perhaps not surprisingly, concur with the relations obtained in (\ref{subeq:range5}) and offer an alternative verification.

Secondly, unlike the time-varying distance function $\bD(t)$, which is infinitely differentiable, $\bB(t)$ is a second-order function under the linear velocity assumption (\ref{eq:constantVelocity}). Differentiating (\ref{eq:dbyy}) yet again, we have \begin{equation}
\label{eq:dbyy_zero} \frac{\ensuremath{d^3} \bB(t)}{\ensuremath{dt^3}}  \Big|_{t=t_0}\
= -0.5\bP (\bR\odot\dddot{\bR} + 3\dot{\bR}\odot\ddot{\bR})\bP=\ \bzero_{N, N},
\end{equation} since generalizing (\ref{eq:rangeL3}) for all $N$ nodes yields
\begin{eqnarray} \frac{d^3 \bR}{dt^3} \triangleq\ \dddot{\bR} = -3\bR^{-1}\odot\dot{\bR}\odot\ddot{\bR}.\end{eqnarray} The result (\ref{eq:dbyy_zero}) is expected, since under the constant velocity assumption \begin{equation}
 \end{equation}

\section{Procrustes alignment} \label{ap:procrustes} Let $\bZ, \underline{\bZ} \in \mathbb{R}^{P \times N}$ matrices which are identical up to a rotation, then
there exists a rotation matrix $\bH$, which minimizes the following cost function
\begin{equation}
\min_{\bH}\ \norm{\bZ- \bH\underline{\bZ}}\quad\ \text{s.t.}\quad \bH^T\bH= \bI_P
\end{equation} and the corresponding optimal \emph{Procrustes rotation} \cite{schonemann1966} is given by
\begin{equation}
\widehat{\bH}=  \bV_z\bU^T_z
\end{equation} where $\bV_z, \bU_z$ are obtained via the singular value decomposition of the matrix product $\underline{\bZ}\bZ^T$, \ie
\begin{equation}
\bU_x\bL_z\bV_z= \underline{\bZ}\bZ^T.
\end{equation}

\section{\Cramer\ Rao Bounds for $\bXu, \bYu$} \label{ap:crbPV}
\subsection{Relative position $\bXu$} The problem of estimating the unknown positions $\bphi_x \triangleq \text{vec}(\bXu) = \begin{bmatrix} \bxu^T_1, \bxu^T_2, \hdots, \bxu^T_{N}\end{bmatrix} ^T \in \mathbb{R}^{NP \times 1}$ from the distance measurements is formulated as \begin{equation}
\ba_x(\bphi_x) -\bd_x= \boldeta_x \label{eq:relXModel}
\end{equation} which is obtained by vectorizing (\ref{eq:range2}).  $\bd_x= [r_{12}, r_{13},\hdots, r_{N(N-1)}] \in \mathbb{R}^{2\barN \times 1}$ is the set of \emph{non-zero} Euclidean distances between $N$ points, with $\barN=\begin{pmatrix} N \\ 2 \end{pmatrix}$. The distance vector is related to the positions by $\ba(\bphi_x) = \begin{bmatrix} a_x(\bxu_1, \bxu_2) , a_x (\bxu_1, \bxu_3),\hdots, a_x (\bxu_{N-1}, \bxu_{N})\end{bmatrix} ^T \in \mathbb{R}^{2\barN \times 1}$ where, \begin{equation}
\ba_x (\bxu_i, \bxu_j) \triangleq\ \big( \bxu^T_i\bxu_i +   \bxu_j^T\bxu_j    -2\bxu^T_i\bxu_j \big)^\frac{1}{2}\ .
\end{equation} Furthermore, the noise plaguing the distance vector is $\boldeta_x \sim \cN(0, \bSigma_{\eta x})$, where $\bSigma_{\eta x}= \text{blkdiag}(\bSigma_r, \bSigma_r)$ and $\bSigma_r$ is given by (\ref{eq:covTheta}).

The \Cramer\ Rao lower Bound (CRB) for any unbiased estimate of $\bphi_x$, is given by the inverse of the Fisher Information Matrix (FIM) \ie \begin{equation} \trace \Big( {\mathbb{E}} \left \{ (\hat{\bphi}_x-\bphi_x)(\hat{\bphi}_x-\bphi_x)^T \right \} \Big) \triangleq  \trace(\bSigma_x) \ge \trace(\bF_x^{-1}) \label{eq:CRBx_}
\end{equation} where $\hat{\bphi}$ is an estimate of the unknown location $\btheta$ and $\bSigma_x$ is the lowest achievable covariance. For the data model (\ref{eq:relXModel}), the FIM $\bF_x \in \mathbb{R}^{NP \times NP}$ is \begin{equation}
\bF_x=
\begin{bmatrix} \dfrac{\partial \ba_x(\bphi_x)}{\partial \bphi^T_x} \end{bmatrix}^T
\bSigma_{\eta y}^{-1}
\begin{bmatrix} \dfrac{\partial \ba_x(\bphi_x)}{\partial \bphi^T_x} \end{bmatrix}\label{eq:FIMx}
\end{equation} where the Jacobian is of the form \begin{equation}
 \dfrac{\partial \ba_x(\bphi_x)}{\partial \bphi^T_x} =
\left[ \dfrac{\partial \ba_x(\bphi_x)}{\partial \bxu^T_1}, \dfrac{\partial \ba_x(\bphi_x)}{\partial \bxu^T_2},\hdots\ ,\dfrac{\partial \ba_x(\bphi_x)}{\partial \bxu^T_N}\right]
\label{eq:Jacobian_x}
\end{equation} whose $i$th element $\left[\dfrac{\partial \ba_x(\bphi)}{\partial \bxu^T_i}\right] $ is given by
\begin{eqnarray}
\left[\dfrac{\partial a(\bxu_1,\bxu_2)^T}{\partial \bxu^T_i}, \dfrac{\partial a(\bxu_1,\bxu_3)^T}{\partial \bxu^T_i}, \hdots, \dfrac{\partial a(\bxu_{N-1},\bxu_{N})^T}{\partial \bxu^T_i} \right] \nonumber
\end{eqnarray} where $\forall 1 \le j,k \le N,\ j\ne k$, we have
\begin{subnumcases}{\dfrac{\partial a(\bxu_j,\bxu_k)}{\partial \bxu^T_i}=}
  d^{-1}_{jk}\left(\bxu_j- \bxu_k\right)^T & if $i = j$  \\
- d^{-1}_{jk}\left(\bxu_j- \bxu_k\right)^T & if $i = k$  \\
  \bzero^T_P.                             & otherwise
\end{subnumcases}

The FIM (\ref{eq:FIMx}) is rank deficient by $3$ for a $P=2$ dimensional scenario \cite{ash2008,chang2006} and is thus non-invertible. Hence, we have the achievable CRB on the relative position as
\begin{equation}
\trace(\bSigma_x) \ge \trace(\bF_x^{\dagger}). \label{eq:CRBx}
\end{equation}

\subsection{Relative velocity $\bYu$} Vectorizing (\ref{eq:ddrange2}), the relative velocity $\bphi_x \triangleq \text{vec}(\bYu) = \begin{bmatrix} \byu^T_1, \byu^T_2, \hdots, \byu^T_{N}\end{bmatrix} ^T \in \mathbb{R}^{NP \times 1}$ estimation is modeled as \begin{equation}
\ba_y(\bphi_y) - \bd^{\odot 2}_y= \bEta_y \label{eq:relYModel}
\end{equation} where $\ba(\bphi_y) = \begin{bmatrix} a_y(\byu_1, \byu_2) , a_y(\byu_1, \byu_3),\hdots, a_y (\byu_{N-1}, \byu_N)\end{bmatrix} ^T$ $\in \mathbb{R}^{2\barN \times 1}$ and \begin{equation}\ba_y (\byu_i, \byu_j) \triangleq\ \byu^T_i\byu_i +   \byu_j^T\byu_j    -2\byu^T_i\byu_j \ . \end{equation} The distance squared vector $\bd^{\odot 2}_y= \{{r}_{ij}{\ddr}_{ij} + {\dr}^{\odot 2}_{ij} \} \forall\ i,j \le N, i \ne j \in \mathbb{R}^{2\barN \times 1}$, where $r_{ij}, \dr_{ij}, \ddr_{ij}$ are the corresponding range estimates. The noise $\bEta_y= \{ \eta_{y,ij} \}$ in the data model is \begin{eqnarray}
\eta_{y,ij} &=&         r_{ij}q_{\ddr,ij} + \ddr_{ij}q_{r,ij} + 2\dr_{ij}q_{\dr,ij} + q_{r,ij}q_{\ddr,ij} + q_{\dr,ij}q_{\dr,ij} \nonumber \\
          &\approx&     r_{ij}q_{\ddr,ij} + \ddr_{ij}q_{r,ij} + 2\dr_{ij}q_{\dr,ij},
\end{eqnarray} where $q_{r,ij}, q_{\dr,ij}, q_{\ddr,ij}$ are the noise variable plaguing the range parameters $r_{ij}, \dr_{ij}, \ddr_{ij}$ respectively. The covariance of the noise is subsequently defined as , \begin{equation}
\bSigma_{\eta y} \    = {\mathbb{E}} \left \{ \bEta_y \bEta^T_y \right \}  \approx\ \text{blkdiag}(\underline{\bSigma}_{\eta y},\underline{\bSigma}_{\eta y})`
\end{equation} where \begin{equation}
\underline{\bSigma}_{\eta y} \    \approx\
\underline{\bR}\bSigma_{\ddr}\underline{\bR} +
\underline{\ddot{\bR}}\bSigma_{r}\underline{\ddot{\bR}} +
4\underline{\dot{\bR}}\bSigma_{\ddr}\underline{\dot{\bR}}\ ,
\end{equation} $\underline{\bR}= \diag(\br), \underline{\dot{\bR}}= \diag(\bdr),\underline{\ddot{\bR}}= \diag(\bddr)$ are the range parameters and $\bSigma_{r}, \bSigma_{\dr},\bSigma_{\ddr}$ are the corresponding covariances matrices (\ref{eq:covTheta}). The \Cramer\ Rao lower Bound (CRB) for $\bphi_y$ is given by  \begin{equation} \trace \Big( {\mathbb{E}} \left \{ (\hat{\bphi}_y-\bphi_y)(\hat{\bphi}_y-\bphi_y)^T \right \} \Big) \triangleq  \trace(\bSigma_y) \ge \trace(\bF_y^{-1}) \label{eq:CRBy_} \end{equation} where $\hat{\bphi_y}$ is an estimate of the unknown velocity $\bphi$ and $\bSigma_y$ is the lowest achievable covariance and $\bF_y \in \mathbb{R}^{NP \times NP}$ is \begin{equation}
\bF_y=
\begin{bmatrix} \dfrac{\partial \ba_y(\bphi_y)}{\partial \bphi^T_y} \end{bmatrix}^T
\bSigma_{\eta y}^{-1}
\begin{bmatrix} \dfrac{\partial \ba_y(\bphi_y)}{\partial \bphi^T_y} \end{bmatrix} \label{eq:FIMy}
\end{equation} where the Jacobian is of the form \begin{equation}
\dfrac{\partial \ba_y(\bphi_y)}{\partial \bphi^T_y} =
\left[ \dfrac{\partial \ba_y(\bphi_y)}{\partial \byu^T_1}, \dfrac{\partial \ba_y(\bphi_y)}{\partial \byu^T_2},\hdots\ ,\dfrac{\partial \ba_y(\bphi_y)}{\partial \byu^T_N}\right]
\label{eq:Jacobian_x}
\end{equation} whose $i$th element $\left[\dfrac{\partial \ba_y(\bphi)}{\partial \byu^T_i}\right] $ is given by
\begin{eqnarray}
\left[\dfrac{\partial a(\byu_1,\byu_2)^T}{\partial \byu^T_i}, \dfrac{\partial a(\byu_1,\byu_3)^T}{\partial \byu^T_i}, \hdots, \dfrac{\partial a(\byu_{N-1},\byu_{N})^T}{\partial \byu^T_i} \right] \nonumber
\end{eqnarray} where $\forall 1 \le j,k \le N,\ j\ne k$, we have
\begin{subnumcases}{\dfrac{\partial a(\byu_j,\byu_k)}{\partial \byu^T_i}=}
2\left(\byu_j- \byu_k\right)^T & if $i = j$  \\
-2\left(\byu_j- \byu_k\right)^T & if $i = k$  \\
  \bzero^T_P.                             & otherwise
\end{subnumcases}

Similar to $\bF_x$, the FIM (\ref{eq:FIMy}) on velocity is also rank degenerate by $3$ for a $P=2$ dimensional case and hence we have the CRB on the relative velocity as \begin{equation}
\trace(\bSigma_y) \ge \trace(\bF_y^{\dagger}). \label{eq:CRBx}
\end{equation}

\bibliographystyle{core/IEEEtran}
\bibliography{IEEEabrv,core/myRef}
\end{document}

%% file: core/macros.tex
\newcommand{\mbf}{\mathbf}

\newcommand{\bzero}{\mbf{0}}

\def\b1{{\mathbf 1}}

\newcommand{\bphi}{\boldsymbol\phi}

\def\bLambda{{\mbox{\boldmath{$\Lambda$}}}}

\def\bphi{{\mbox{\boldmath{$\phi$}}}}

\def\btau{{\mbox{\boldmath{$\tau$}}}}

\def\bSigma{{\mbox{\boldmath{$\Sigma$}}}}

\newcommand{\cN}{\ensuremath{\mathcal{N}}}

\newcommand{\bXu}{\ensuremath{\underline{\bX}}}
\newcommand{\bYu}{\ensuremath{\underline{\bY}}}

\newcommand{\argmin}{\mbox{\rm arg}\min}

\newcommand{\diag}{\mbox{\rm diag}}
\newcommand{\bdiag}{\mbox{\rm bdiag}}

\def\eg{{e.g.,\ }}
\def\ie{{i.e.,\ }}

\newcommand{\betab}{\begin{tabbing}}
\newcommand{\entab}{\end{tabbing}}
\newcommand{\beitem}{\begin{itemize}}
\newcommand{\enitem}{\end{itemize}}
\newcommand{\bea}{\begin{array}}
\newcommand{\ena}{\end{array}}
\newcommand{\beq}{\begin{equation}}
\newcommand{\enq}{\end{equation}}
\newcommand{\beqa}{\begin{eqnarray}}
\newcommand{\enqa}{\end{eqnarray}}
\newcommand{\beqan}{\begin{eqnarray*}}
\newcommand{\enqan}{\end{eqnarray*}}
\newcommand{\beenum}{\begin{enumerate}}
\newcommand{\enenum}{\end{enumerate}}
\newcommand{\DL}{\begin{dashlist}}
\newcommand{\DLE}{\end{dashlist}}

\newcommand{\ba}{{\ensuremath{\mathbf{a}}}}
\newcommand{\bb}{{\ensuremath{\mathbf{b}}}}

\newcommand{\bd}{{\ensuremath{\mathbf{d}}}}
\newcommand{\be}{{\ensuremath{\mathbf{e}}}}
\newcommand{\bff}{{\ensuremath{\mathbf{f}}}}
\newcommand{\bg}{{\ensuremath{\mathbf{g}}}}
\newcommand{\bh}{{\ensuremath{\mathbf{h}}}}

\newcommand{\bq}{{\ensuremath{\mathbf{q}}}}
\newcommand{\br}{{\ensuremath{\mathbf{r}}}}

\newcommand{\bt}{{\ensuremath{\mathbf{t}}}}

\newcommand{\bx}{{\ensuremath{\mathbf{x}}}}
\newcommand{\by}{{\ensuremath{\mathbf{y}}}}
\newcommand{\bz}{{\ensuremath{\mathbf{z}}}}

\newcommand{\bA}{{\ensuremath{\mathbf{A}}}}
\newcommand{\bB}{{\ensuremath{\mathbf{B}}}}

\newcommand{\bD}{{\ensuremath{\mathbf{D}}}}

\newcommand{\bF}{{\ensuremath{\mathbf{F}}}}
\newcommand{\bG}{{\ensuremath{\mathbf{G}}}}
\newcommand{\bH}{{\ensuremath{\mathbf{H}}}}
\newcommand{\bI}{{\ensuremath{\mathbf{I}}}}
\newcommand{\bJ}{{\ensuremath{\mathbf{J}}}}

\newcommand{\bL}{{\ensuremath{\mathbf{L}}}}

\newcommand{\bP}{{\ensuremath{\mathbf{P}}}}

\newcommand{\bR}{{\ensuremath{\mathbf{R}}}}

\newcommand{\bT}{{\ensuremath{\mathbf{T}}}}
\newcommand{\bU}{{\ensuremath{\mathbf{U}}}}
\newcommand{\bV}{{\ensuremath{\mathbf{V}}}}

\newcommand{\bX}{{\ensuremath{\mathbf{X}}}}
\newcommand{\bY}{{\ensuremath{\mathbf{Y}}}}
\newcommand{\bZ}{{\ensuremath{\mathbf{Z}}}}

\def\bSigma{{\mbox{\boldmath{$\Sigma$}}}}

\def\btheta{{\mbox{\boldmath{$\theta$}}}}

\def\btau{{\mbox{\boldmath{$\tau$}}}}

\def\bEta{{\mbox{\boldmath{$\eta$}}}}

\def\wrt{{w.r.t.\ }}
\def\iid{{i.i.d.\ }}
\newcommand{\norm}[1]{\lVert#1\rVert}

\def\ur{{\underline{r}}}
\def\bur{{\underline{\br}}}

\newcommand{\Cramer}{Cram\'{e}r}

\newcommand{\barN}{{\ensuremath{\bar{N}}}}

\newcommand{\bxu}{{\ensuremath{\underline{\bx}}}}
\newcommand{\byu}{{\ensuremath{\underline{\by}}}}

\newcommand{\thetau}{{\ensuremath{\underline{\theta}}}}
\newcommand{\bthetau}{{\ensuremath{\underline{\btheta}}}}
\newcommand{\trace}{{\ensuremath{\text{Tr}}}}

\def\boldeta{{\mbox{\boldmath{$\eta$}}}}